\documentclass[12pt]{article}

\usepackage{amsmath}
\usepackage{graphicx}
\usepackage{amsfonts}
\usepackage{amssymb}
\newtheorem{defin}{\sc {Definition}}[section]
\newtheorem{definition}[defin]{\sc {Definition}}
\newtheorem{theorem}[defin]{\sc {Theorem}}
\newtheorem{lemma}[defin]{\sc {Lemma}}
\newtheorem{corollary}[defin]{\sc {Corollary}}

\newtheorem{proposition}[defin]{\sc {Proposition}}

\newtheorem{remark}[defin]{\sc {Remark}}
\newtheorem{example}[defin]{\sc {Example}}

\newcommand{\numbercellong}[2]
{
\begin{picture}(80,20)(0,0)
\put(0,0){\framebox(80,20)} \put(40,10){\makebox(0,0){#1}}
\end{picture}
}

\begin{document}
\title{Cooperation under Incomplete Information on the Discount Factors}
\author{C Maor}

\begin{titlepage}
\begin{center}


\LARGE
\textbf{Cooperation under Incomplete Information on the Discount Factors}\\
\vspace{1.5cm}
\large
Thesis submitted in partial fulfillment of requirements for the M. Sc. degree in the School of Mathematical Sciences, Tel-Aviv University\\
\vspace{0.2cm}
by\\
\vspace{0.2cm}
\Large
C Maor\\
\vspace{2cm}
\large
The research work for this thesis has been carried out at Tel-Aviv University under the supervision of Prof. Eilon Solan\\
\vspace{0.5cm}
\Large
August 2010
\normalsize
\end{center}

\end{titlepage}

\newpage
\section* {Acknowledgments}
I would like to thank my advisor, professor Eilon Solan, for the time, effort and goodwill he invested helping me with my research. I would also like to thank professor Dov Samet and proffesor Ehud Lehrer, whose remarks helped to improve this work. Many thanks to my fellow graduate students Roee Teper, Yuval Heller and Ya'arit Even. Last but not least, I thank Yuval Elhanati for his support and good advices.

\newpage
\begin{abstract}
In the repeated Prisoner's Dilemma, when every player has a different discount factor, the grim-trigger strategy is an equilibrium if and only if the discount factor of each player is higher than some threshold. What happens if the players have incomplete information regarding the discount factors? In this work we look at repeated games in which each player has incomplete information regarding the other player's discount factor, and ask when a pair of grim-trigger strategies is an equilibrium. We provide necessary and sufficient conditions for such strategies to be an equilibrium. We characterize the states of the world in which the strategies are not triggered, i.e., the players cooperate, in such equilibria (or $\epsilon$-equilibria), and ask whether these ``cooperation events'' are close to those in the complete information case, when the information is ``almost'' complete, in several senses.
\end{abstract}

\newpage
\tableofcontents

\newpage
\section{Introduction}

In the repeated Prisoner's Dilemma, when every player has a different discount factor, the grim-trigger strategy is an equilibrium if and only if the discount factor of each player is sufficiently close to 1. A similar situation holds for other repeated two-player games in which there is a pair of pure actions $\tau=(\tau_1,\tau_2)$ such that under $\tau$ the payoff for each player is strictly higher than some equilibrium payoff: there are two thresholds $\lambda_1^0,\lambda_2^0$ such that the grim-trigger course of action $\tau^*=(\tau_1^*,\tau_2^*)$, under which the players follows $\tau$ until a deviation occur, and then they switch to the equilibrium action that punishes the other player, is an equilibrium if and only if $\lambda_i\ge\lambda_i^0$ for $i=1,2$. In a symmetric game, like the Prisoner's Dilemma, these thresholds are the same for both players.\footnote{Note that when the players have different discount factors, there may exist equilibria which yield payoffs that are higher than the payoffs under any such $\tau^*$. Such cooperative equilibria are not in the scope of this work. See Lehrer and Pauzner (1999).}

In this work we look at repeated games in which each player has incomplete information regarding the other player's discount factor, and ask when a pair of grim-trigger strategies is an equilibrium. A strategy in the incomplete information game is information-dependent: a strategy assigns a ``course of action'', which is a strategy in the repeated game (with complete information), to each state of the world.
A pair of strategies will be called ``conditional-grim-trigger'' if it is composed of two action pairs $\sigma=(\sigma_1,\sigma_2)$ and $\tau=(\tau_1,\tau_2)$, where (a) $\sigma$ is an equilibrium of the one-shot game, (b) the payoff under $\tau$ is higher than the payoff under $\sigma$ for both players, and (c) in any state of the world $\omega$, player $i$ either plays repeatedly $\sigma_i$, or plays $\tau_i$ until a deviation from $\tau$ is detected, and then he switches to playing $\sigma_i$ forever, for $i=1,2$.
\footnote{This is a narrower sense of the concept of grim-trigger strategy, since we demand that the pair of punishing strategies will define an equilibrium, instead of any pair $\sigma=(\sigma_1,\sigma_2)$ under which $\max_{\sigma_i'}u_i(\sigma_i',\sigma_j)<u_i(\tau)$ (Here and below, $i$ is an arbitrary player and $j$ is the player which is not $i$). Because of this assumption, a conditional-grim-trigger strategy pair Bayesian equilibrium is a perfect Bayesian equilibrium.}

In order for a player to cooperate (assuming his own discount factor is sufficiently high), he needs to ascribe high enough probability that the other player's discount factor is higher than his threshold. But he also needs to ascribe high enough probability that the other player ascribes high enough probability that his own discount factor is higher than his own threshold, and so on. We thus get that infinitely many conditions need to hold in order for the conditional-grim-trigger strategy pair to be an equilibrium, one for each level of belief for each player. Note that the ``high enough'' probability in each level depends on the player's own discount factor, since the higher his discount factor is, the player will lose more if the grim-trigger course of action is triggered. Therefore, a player with high discount factor will cooperate in situations where he wouldn't have cooperated if his discount factor was lower.

We show that this is \emph{not} the case, and only two conditions for each player are necessary and sufficient to characterize when the conditional-grim-trigger strategy is a Bayesian equilibrium. In this strategy, each player plays the grim-trigger course of action in some states of the world, which is called his ``cooperation event'', and the punishing strategy in the others. We show that this strategy is a Bayesian equilibrium if and only if (a) each player plays the grim-trigger course of action only when his discount factor is above his threshold; and (b) each player ascribes sufficiently high probability to the other player's cooperation event whenever he plays the grim-trigger course of action, and a sufficiently low probability to the other player's cooperation event whenever he does not play this course of action. This result holds for all belief structures, whether they are derived from a common prior or not.

We describe the sets of states of the world that satisfy these conditions, and relate them to the concept of $f$-belief and common-$f$-belief. These two concepts generalize the concepts of $p$-belief and common-$p$-belief defined by Samet and Mondrer (1989).
In particular, we show that for the repeated Prisoner's Dilemma, the strategy profile above is an equilibrium whenever each player plays the cooperation strategy if he $f$-believes that it is a common-$f$-belief that both players' discount factors are above a given threshold. For games with more then two actions for each player, an additional condition is needed.

We also show that these conditions are sufficient, though not necessary, for this kind of strategy to be a Bayesian equilibrium even if each player does not know his own discount factor, and also in a larger class of two-player games with incomplete information, in which there is an equilibrium which holds in all states of the nature (equivalent to the ``punishment" equilibrium) and an equilibrium which holds only in some states of nature (equivalent to the ``cooperation" grim-trigger equilibrium, which holds only for high discount factors). We also show that similar conditions are sufficient, though not necessary, in repeated games with incomplete information with more than two players.

Last, we look at belief spaces in which the information is ``almost'' complete, in several senses, and see that in some senses, when the information is almost complete, there is a conditional-grim-trigger $\epsilon$-equilibrium in which the players cooperate in all states of the world where they could have cooperated under equilibrium in the complete information case, but for a set of small probability, while in other senses, there may be no conditional-grim-trigger $\epsilon$-equilibrium at all.

\section{The Model}
\label{section model}

Let $\Gamma=\left(N,(A_i)_{i\in N},(u_i)_{i\in N}\right)$ be a two-player one-shot game: $N=\{1,2\}$ is the set of players, and for every player $i\in N$, $A_i$ is the set of pure actions of player $i$, and $u_i$ is his utility function (extended multilinearly to mixed strategies).
Let $\sigma=(\sigma_1,\sigma_2)$ be a Nash equilibrium in mixed-strategies in $\Gamma$.
Assume that the payments in another non-equilibrium pure action profile $\tau=(\tau_1,\tau_2)$,
which we call a \emph{cooperation profile}, are higher than the equilibrium payments,
that is $u_i(\tau)>u_i(\sigma)$ for $i=1,2$.
Also, assume that for $i=1,2$, $\tau_i$ is \emph{not} a best response to $\sigma_j$, and, in particular, not in the support\footnote{We discuss this assumption in \ref{supportremark} below.} of $\sigma_i$.

Let $G=\left(N,(S,\mathcal{S}),\Pi,(A_i)_{i\in N},(u_i)_{i\in N}\right)$ be the repeated game based on $\Gamma$, with incomplete information regarding the discount factors:
\begin{itemize}
  \item $N=\{1,2\}$ is the set of players.
  \item $(S,\mathcal{S})$ is a measurable space of the states of nature, that is, $S\subseteq[0,1)^2$ is the set of possible pairs of discount factors of the players. $\mathcal{S}$ is the $\sigma$-algebra that is induced on $S$ by the Borel $\sigma$-algebra on $[0,1)^2$.
  \item $\Pi=(\Omega,\Sigma,\lambda,(P_i)_{i\in N})$ is the players' belief space:
  \begin{itemize}
    \item $(\Omega,\Sigma)$ is a measurable space of states of the world.
    \item $\lambda:\Omega\rightarrow S$ is a measurable function between the states of the world and the states of nature, i.e., the players' discount factors are $\lambda(\omega)=(\lambda_1(\omega),\lambda_2(\omega))$,
        where $\lambda_i(\omega)$ is player $i$'s discount factor in the state of the world $\omega$. The definition of $\mathcal{S}$ implies that $\{\omega\mid\lambda(\omega)\in B\}\in\Sigma$, for every open $B\subseteq[0,1)^2$.
    \item $P_i:\Omega\rightarrow\Delta(\Omega)$ is a measurable function that assigns a belief for player $i$ to each state of the world $\omega$. We denote by $P_i(E\mid\omega)$ the probability that player $i$ ascribes to the event $E\subseteq\Omega$ at the state of the world $\omega$, and by $E_i(\cdot\mid\omega)$ the corresponding expectation operator. $P_i$ is measurable in the sense that for every $E\subseteq\Omega$, $P_i(E\mid\cdot)$ is a measurable function. $P_i$ is consistent, in the sense that each player knows his belief: $P_i(\{\omega':P_i(\omega)=P_i(\omega')\}\mid\omega)=1$, for every $\omega\in\Omega$. We assume that $P_i$ is such that each player knows his own discount factor in every state of the world $\omega$: $P_i(\{\omega':\lambda_i(\omega')=\lambda_i(\omega)\}\mid\omega)=1$ for every $\omega\in\Omega$. In Section \ref{secnoselfinf} we drop this assumption.
  \end{itemize}
  Because $P_i$ is consistent, it divides $\Omega$ into disjoint ``types'' of player $i$; that is, into the sets $\big\{\{\omega'\mid P_i(\omega')=P_i(\omega)\},\omega\in\Omega\big\}$. Denote by $\Sigma_i\subseteq\Sigma$ the $\sigma$-algebra generated by these sets, and by the sets $\{\omega\mid\lambda_i(\omega)\in B\}$, for every open set $B\subseteq[0,1)$ (since each player knows his own discount factor, these sets are unions of his types). We will call a subset of $\Omega$ or a function from $\Omega$ \emph{$i$-measurable} if it is measurable with respect to $\Sigma_i$.
  \item $A_i$ and $u_i$ are the same as in $\Gamma$: the set of pure actions of player $i$, and his utility function (in each stage of the game), which are independent of the state of the world.
\end{itemize}

A \emph{course of action} of player $i$ is function that assigns a mixed action of player $i$ to each finite history of actions in the game. This function is independent of the state of the world.
A \emph{strategy} of player $i$ is an $i$-measurable function $\eta_i$ that assigns a course of action $\eta_i(\omega)$ to each state of the world $\omega$.
The payoff of player $i$ when the profile $(\eta_i,\eta_j)$ is played, conditional on the state of the world $\omega$, is $\gamma_i(\eta_i,\eta_j\mid\omega)=E_i(\sum_{t=1}^\infty\lambda_i(\omega)^{t-1}u_i^t\mid\omega)$, where $u_i^t$ is the utility in stage $t$. The expectation\footnote{Note that this is actually the expected subjective payoff of player $i$, because the expectation is taken with respect to \emph{his} belief $P_i(\omega)$. This is the relevant payoff for the player's decision making. One could define the payoff as $\gamma_i=E_i((1-\lambda_i(\omega))\sum_{t=1}^\infty\lambda_i(\omega)^{t-1}u_i^t\mid\omega)$. Most of our results remain unchanged with this payoff function.} depends on $\omega$, since player $j$'s actions and player $i$'s belief may depend on $\omega$. As mentioned, we assume that the discount factor $\lambda_i$ is known to player $i$.

Recall that $\tau=(\tau_1,\tau_2)$ is a cooperative profile, and $\sigma=(\sigma_1,\sigma_2)$ is a Nash equilibrium with payoffs lower than those under $\tau$.
\begin{definition}
\emph{A grim-trigger course of action} for player $i$, based on $\sigma$ and $\tau$, is the course of action $\tau_i^*$ under which player $i$ plays $\tau_i$ until player $j$ deviates from $\tau_j$, and from that stage on plays $\sigma_i$.
\end{definition}
In the complete information case, there are thresholds $\lambda_1^0,\lambda_2^0$ such that $\tau^*$ is an equilibrium if and only if $\lambda_i\ge \lambda_i^0$:
$$\lambda_i^0:=\min\left\{\lambda_i\mid
\frac{u_i(\tau)}{1-\lambda_i}-(u_i(\sigma_i',\tau_j)+u_i(\sigma)\frac{\lambda_i}{1-\lambda_i})\ge0 \;\;\forall \sigma_i'\ne\tau_i\right\}.$$
Denote $\Lambda_i=\{\omega\in\Omega\mid\lambda_i(\omega)\ge\lambda_i^0\}$,
which is the set of states in which player $i$ cannot profit by deviating from the profile $\tau^*$. Note that player $i$ always knows whether $\lambda_i(\omega)\ge\lambda_i^0$ or not: $\Lambda_i$ is an $i$-measurable event.

We are interested only in ``conditional-grim-trigger'' strategies:
\begin{definition}
\label{grimdef}
A strategy $\eta_i$ of player $i$ is called \emph{conditional-grim-trigger strategy} (with respect to $\sigma$ and $\tau$) if there is an $i$-measurable set $K_i\subseteq\Omega$ such that:
$$
\eta_i(\omega)=
\left\{\begin{array}{ll}
\tau_i^*&\omega\in K_i\\
\sigma_i^*&\omega\notin K_i
\end{array}
\right.
$$
Here $\sigma_i^*$ is the course of action where player $i$ always plays $\sigma_i$. We denote this strategy by $\eta_i^*(K_i)$.
\end{definition}

Note that if a pair of grim-trigger strategies $\eta^*(K_1,K_2)=(\eta_1^*(K_1),\eta_2^*(K_2))$ is played, then, because $\tau_i$ is not in $\sigma_i$'s support, after the first stage both players learn if the other player ``cooperates''. From that point on they do not learn anything else.

\begin{definition}
If $\eta^*(K_1,K_2)$ is a Bayesian equilibrium, the pair of events $(K_1,K_2)$ is called \emph{cooperation events}.
\end{definition}
The existence of non-empty cooperation events guarantees that the players may cooperate in some states of the world.

Note that $\eta^*(\emptyset,\emptyset)$ is a Bayesian equilibrium in which the players always follow $\sigma$. Also note that if $\eta^*(K_1,K_2)$ is a Bayesian equilibrium and $K_i=\emptyset$, then $K_j=\emptyset$, since we assumed that $\tau_j$ is not a best response to $\sigma_i$.

\section{Main Result: Characterization of the Cooperation Events}

\subsection{The Complete Information Case}

When the game has complete information we can identify $S$ and $\Omega$, and each player knows the true state of the world.

\begin{theorem}
When the game $G$ has complete information, $\eta^*(K_1,K_2)$ is a Bayesian equilibrium if and only if $K_1=K_2\subseteq\Lambda$ where $\Lambda=\Lambda_1\cap\Lambda_2=\{\lambda_k\ge\lambda_k^0\:\mbox{for $k=1,2$}\}$.
\end{theorem}

\noindent\textbf{Proof: } Assume $K_1=K_2\subseteq\Lambda$. Then for every $\omega\in K_1$, the profile $\tau^*$ is played, and both players know it. Since $\omega\in\Lambda$, both $\lambda_1\ge\lambda_1^0$ and $\lambda_2\ge\lambda_2^0$, so neither player can profit by deviating. For every $\omega\notin K_1$, the profile $\sigma^*$ is played, and both players know it. Since $\sigma$ is a Nash equilibrium in $\Gamma$, neither player can profit by deviating. For the opposite direction, there are two cases: $K_1=K_2\not\subset\Lambda$ and $K_1\ne K_2$.
Assume first that $K_1=K_2\not\subset\Lambda$, and let $\omega\in K_1/\Lambda$. Without loss of generality assume that $\lambda_1<\lambda_1^0$ in the state of the world $\omega$. In the state of the world $\omega$ the profile $\tau^*$ is played, but since $\lambda_1<\lambda_1^0$, player 1 can profit by deviating, so $\eta^*(K_1,K_2)$ is not a Bayesian equilibrium. Assume Now that $\omega\in K_1/K_2$. Then in the state of the world $\omega$, player 1 knows that player 2 plays $\sigma_2^*$, and since $\tau_1$ is not a best response to $\sigma_2$, it is profitable for him to deviate from $\tau_1^*$ to $\sigma_1^*$, and therefore $\eta^*(K_1,K_2)$ is not a Bayesian equilibrium. $\Box$

\begin{example}
\label{prisonex}
Prisoner's Dilemma with complete information.
\newline \begin{picture}(210,80)(-80,0)
\put( 15,8){$C$}
\put( 15,28){$D$}
\put( 80,50){$D$}
\put(160,50){$C$}
\put( 40, 0){\numbercellong{$0,4$}{}}
\put(40,20){\numbercellong{$1,1$}{}}
\put(120,0){\numbercellong{$3,3$}{}}
\put(120,20){\numbercellong{$4,0$}{}}
\end{picture}
\newline
Here $\sigma=(D,D)$ and $\tau=(C,C)$. It can be easily calculated that $\lambda_1^0=\lambda_2^0=1/3$, and that $\eta^*(K_1,K_2)$ is a Bayesian equilibrium if and only if $K_1=K_2\subseteq\{\lambda_1,\lambda_2\ge1/3\}$.
\end{example}

\subsection{The Incomplete Information Case}

Our main result is the following:

\begin{theorem}
\label{conditionprop}
Let $K_i\subseteq\Omega$ be an $i$ measurable event for $i=1,2$, then the strategy profile $\eta^*(K_1,K_2)=(\eta_1^*(K_1),\eta_2^*(K_2))$ is a Bayesian equilibrium if and only if ,for $i=1,2$, $K_i\subseteq\Lambda_i$ and
\begin{enumerate}
  \item $P_i(K_j\mid\omega)\ge f_i(\omega)$ for every $\omega\in K_i$,
  \item $P_i(K_j\mid\omega)\le g_i(\omega)$ for every $\omega\notin K_i$,
\end{enumerate}
for the $i$-measurable functions\footnote{By definition, the infimum over an empty set is 1 and the supremum over an empty set is 0.}
$$
\begin{array}{ll}
      f_i(\omega):=
      \max_{\sigma_i'\in F_i}
      \frac{u_i(\sigma_i',\sigma_j)-u_i(\tau_i,\sigma_j)}{\left(\frac{u_i(\tau)}{1-\lambda_i(\omega)}-(u_i(\sigma_i',\tau_j)
      +u_i(\sigma)\frac{\lambda_i(\omega)}{1-\lambda_i(\omega)})\right)+(u_i(\sigma_i',\sigma_j)-u_i(\tau_i,\sigma_j))},
\end{array}
$$
and
$$
g_i(\omega):=\min\{g_i^1,g_i^2(\omega),g_i^3(\omega)\},
$$
where $\sigma_i'$ is an action of player $i$, and
$F_i=\{\sigma_i'\mid u_i(\tau_i,\sigma_j)<u_i(\sigma_i',\sigma_j)\}$.
$g_i^1$, $g_i^2(\omega)$ and $g_i^3(\omega)$ are derived from different kinds of deviations, and are defined by:
$$g_i^1:=\min_{\sigma_i'\in H_i^1}
        \frac{u_i(\sigma)-u_i(\sigma_i',\sigma_j)}
        {(u_i(\sigma)-u_i(\sigma_i',\sigma_j))+(u_i(\sigma_i',\tau_j)-u_i(\sigma_i,\tau_j))}$$
where $H_i^1:=\left\{\sigma_i'\ne\sigma_i,\tau_i\mid u_i(\sigma_i,\tau_j)<u_i(\sigma_i',\tau_j)\right\}$,
$$
        g_i^2(\omega):=
        \frac{u_i(\sigma)-u_i(\tau_i,\sigma_j)}{(u_i(\sigma)-u_i(\tau_i,\sigma_j))+
        \left(\frac{u_i(\tau)}{1-\lambda_i(\omega)}-(u_i(\sigma_i,\tau_j)
        +u_i(\sigma)\frac{\lambda_i(\omega)}{1-\lambda_i(\omega)})\right)}
$$
whenever $u_i(\sigma_i,\tau_j)+u_i(\sigma)\left(\frac{\lambda_i(\omega)}{1-\lambda_i(\omega)}\right)
<\left(\frac{u_i(\tau)}{1-\lambda_i(\omega)}\right)$, and $g_i^2(\omega):=1$ otherwise, and
$$
\begin{array}{l}
        g_i^3(\omega):=
        \min_{\sigma_i'\in H_i^3(\omega)}
        \frac{u_i(\sigma)-u_i(\tau_i,\sigma_j)}
        {(u_i(\sigma)-u_i(\tau_i,\sigma_j))+(u_i(\tau)-u_i(\sigma_i,\tau_j)+
        (u_i(\sigma_i',\tau_j)-(u_i(\sigma)))(\lambda_i(\omega))},
\end{array}
        $$
where
$$
        H_i^3(\omega):=\left\{\sigma_i'\ne\tau_i\mid
        u_i(\tau)-u_i(\sigma_i,\tau_j)+(u_i(\sigma_i',\tau_j)-(u_i(\sigma)))\lambda_i(\omega)<0\right\}.
$$
\end{theorem}

We now present an equivalent formulation of Theorem \ref{conditionprop}, using the concept of $f$-belief and common-$f$-belief, which we define now.
\begin{definition}
\label{definition fbelief}
Let $N$ be a set of players and $(\Omega,\Sigma,\lambda,(P_i)_{i\in N})$ be a general belief space on a measurable
space of states of nature $(S,\mathcal{S})$. Let $f:\Omega^N \rightarrow \mathbb{R}^N$ be a measurable function, where $f_i$ is $i$-measurable for every $i\in N$. Let $A\subseteq\Omega$ be an event. We say that \emph{player $i$ $f$-believes in the event $A$ at $\omega$} if $P_i(A\mid\omega)\ge f_i(\omega)$. We say that \emph{the event $A$ is an common-$f$-belief at $\omega$} if in the state of the world $\omega$ each player $f$-believes in $A$, $f$-believes that each other player $f$-believes in $A$, $f$-believes that each other player $f$-believes that each player $f$-believes in $A$, etc.
\end{definition}
These two concepts generalize the concepts of $p$-belief and common-$p$-belief described by Samet and Mondrer (1989). These concepts are discussed in detail in Section \ref{fbeliefsec}.

The definitions imply that the conditions in Theorem \ref{conditionprop} are equivalent to the following:
\begin{enumerate}
  \item $K_j$ is an $f$-belief in $K_i$.
  \item $K_j^c$ is an $(1-g)$-belief in $K_i^c$.
\end{enumerate}
In Section \ref{fbeliefsec} we prove that these conditions are also equivalent to the following condition: each player either
$f$-believes that $K_1\cap K_2$ is a common-$f$-belief or
$(1-g)$-believes that the event ``$K_1\cap K_2$ is not a common-$f$-belief'' is a common-$(1-g)$-belief (Theorem \ref{theorem1}).

\subsection{Proof of Theorem \ref{conditionprop}}

Player $i$'s payoff under the strategy profile $\tau^*$ is $\gamma_i(\tau^*)=\frac{u_i(\tau)}{1-\lambda_i(\omega)}$.
As mentioned before, the strategy profile $\tau$ is not an equilibrium in the one-shot game, whereas $\sigma$ is.
If player $i$ deviates from the profile $\tau^*$ and plays $\sigma_i'\ne\tau_i$ in the first stage, then from the second stage on player $j$ will play $\sigma_j$, to which $\sigma_i$ will be player $i$'s best response.
The expected payoff for player $i$ will then be $u_i(\sigma_i',\tau_j)+u_i(\sigma)\frac{\lambda_i(\omega)}{1-\lambda_i(\omega)}$. If there is a profitable deviation from $\tau^*$ for player $i$, there is such a profitable deviation in the first stage.
Therefore $\lambda_i^0:=\min\left\{\lambda_i\mid
\frac{u_i(\tau)}{1-\lambda_i}-(u_i(\sigma_i',\tau_j)+u_i(\sigma)\frac{\lambda_i}{1-\lambda_i})\ge0
\;\;\forall \sigma_i'\ne\tau_i\right\}$, is the minimal discount factor $\lambda_i$ such that player $i$ cannot gain by deviating from the profile $\tau^*$.

\noindent We will now check player $i$'s options to deviate from the profile $\eta^*(K_1,K_2)$.
\newline
\textbf{Case 1:} $\omega\in K_i$.

Note that, because $K_i$ is $i$-measurable, in this case $P_i(K_i\mid\omega)=1$.

Player $i$'s payoff under the strategy profile $\eta^*(K_i,K_j)$ is
$$
\begin{array}{ll}
\gamma_i(\eta^*(K_i,K_j)\mid\omega)=
\\
\;\;\;\;P_i(K_j\mid\omega)\left(\frac{u_i(\tau)}{1-\lambda_i(\omega)}\right)
+(1-P_i(K_j\mid\omega))\left(u_i(\tau_i,\sigma_j)+u_i(\sigma)\frac{\lambda_i(\omega)}{1-\lambda_i(\omega)}\right).
\end{array}
$$
Indeed, according to player $i$'s belief, with probability $P_i(K_j\mid\omega)$ player $j$ plays $\tau_j^*$, so they will play $\tau$ at every stage and his payoff will be $\frac{u_i(\tau)}{1-\lambda_i(\omega)}$, and with probability $(1-P_i(K_j\mid\omega))$ player $j$ plays $\sigma_j^*$ so in the first stage the profile played will be $(\tau_i,\sigma_j)$ and afterwards the players will play $\sigma$, so that player $i$'s payoff will be $u_i(\tau_i,\sigma_j)+u_i(\sigma)\frac{\lambda_i(\omega)}{1-\lambda_i(\omega)}$.

We now check the conditions that guarantee that player $i$ cannot profit by deviating:
\begin{itemize}
  \item Because $\sigma$ is an equilibrium, deviation after the first stage can be profitable for player $i$
        only if player $j$ played $\tau_j$ in the first stage.
        Also, in that case, if it's profitable to deviate in stage $k>2$, it is also profitable to deviate in stage $k=2$, because when $\eta^*(K_1,K_2)$ is played, the players do not learn anything from the second stage onwards.
        After the first deviation the best response to $\eta_j^*(K_j)$ is to play $\sigma_i$ in all stages, because player $j$ will play $\sigma_j$.
        Denote by $\sigma_i'^{**}$ the course of action in which player $i$ plays $\tau_i$ in the first stage, an if player $j$ played $\tau_j$ in the first stage, player $i$ plays a pure action $\sigma_i'\ne\tau_i$ in stage $2$ and $\sigma_i$ afterwards. If player $j$ played $\sigma_j$ in the first stage, player $i$ plays $\sigma_i$ from the second stage onwards. The payoff is:
$$
\begin{array}{ll}
        \gamma_i(\sigma_i'^{**},\eta_j^*(K_j)\mid\omega)=\\
        \;\;\;\;=P_i(K_j\mid\omega)\left(u_i(\tau)+u_i(\sigma_i',\tau_j)\lambda_i(\omega)+
        u_i(\sigma)\frac{\lambda_i^2(\omega)}{1-\lambda_i(\omega)}\right)+\\
        \;\;\;\;\;\;\;\;+(1-P_i(K_j\mid\omega))\left(u_i(\tau_i,\sigma_j)+u_i(\sigma)\frac{\lambda_i(\omega)}{1-\lambda_i(\omega)}\right).
\end{array}
$$
        So that $\eta^*(K_1,K_2)$ is an equilibrium we should have $\gamma_i(\eta^*(K_1,K_2)\mid\omega)\ge\gamma_i(\sigma_i'^{**},\eta_j^*(K_j)\mid\omega)$ for all $\sigma_i'\ne\tau_i$, and therefore,
        $$
        P_i(K_j\mid\omega)\lambda_i(\omega)\left(
        \frac{u_i(\tau)}{1-\lambda_i(\omega)}-(u_i(\sigma_i',\tau_j)+u_i(\sigma)\frac{\lambda_i(\omega)}{1-\lambda_i(\omega)})
        \right)\ge 0,
        $$
        for every $\sigma_i'\ne\tau_i$.
        Because $\lambda_i>0$, either $P_i(K_j\mid\omega)=0$ or $\omega\in\Lambda_i$.
  \item Player $i$ can also deviate in the first stage to a pure strategy $\sigma_i'\ne\tau_i$.
        As before, after the deviation his best response to $\eta_j^*$ is to play $\sigma_i$ in all stages.
        His payoff in this case is (denote by $\sigma_i^*$ the course of action of player $i$ that plays $\sigma_i'\ne\tau_i$ at stage 1 and $\sigma_i$ thereafter):
\begin{eqnarray}
      \gamma_i(\sigma_i'^*,\eta_j^*(K_j)\mid\omega)&=&
      P_i(K_j\mid\omega)\left(u_i(\sigma_i',\tau_j)+u_i(\sigma)\frac{\lambda_i(\omega)}{1-\lambda_i(\omega)}\right)+
\nonumber\\
      &+&(1-P_i(K_j\mid\omega))\left(u_i(\sigma_i',\sigma_j)+u_i(\sigma)\frac{\lambda_i(\omega)}{1-\lambda_i(\omega)}\right).
\nonumber
\end{eqnarray}
      In order for $\eta^*(K_1,K_2)$ to be an equilibrium we should have

\noindent $\gamma_i(\eta^*(K_1,K_2)\mid\omega)\ge\gamma_i(\sigma_i'^*,\eta_j^*(K_j)\mid\omega)$, and therefore, for every $\sigma_i'\ne\tau_i$:
\begin{eqnarray}
\label{eq1}
      P_i(K_j\mid\omega)\left(\frac{u_i(\tau)}{1-\lambda_i(\omega)}-(u_i(\sigma_i',\tau_j)
      +u_i(\sigma)\frac{\lambda_i(\omega)}{1-\lambda_i(\omega)})\right)+
\nonumber\\
      +(1-P_i(K_j\mid\omega))(u_i(\tau_i,\sigma_j)-u_i(\sigma_i',\sigma_j))\ge0.
\end{eqnarray}
      Because $\tau_i$ is not a best response to $\sigma_j$, $P_i(K_j\mid\omega)>0$, otherwise inequality (\ref{eq1}) becomes $u_i(\tau_i,\sigma_j)-u_i(\sigma_i',\sigma_j)\ge0$ for every $\sigma_i'\ne\tau_i$ in contradiction. As above we deduce, that $\omega\in\Lambda_i$, or $K_i\subseteq\Lambda_i$. Therefore we get that
      $\frac{u_i(\tau)}{1-\lambda_i(\omega)}-(u_i(\sigma_i',\tau_j)
      +u_i(\sigma)\frac{\lambda_i(\omega)}{1-\lambda_i(\omega)})\ge0$.
      Therefore, if $u_i(\tau_i,\sigma_j)\ge u_i(\sigma_i',\sigma_j)$ inequality \ref{eq1} trivially holds.
      Otherwise we obtain:
      $$
\begin{array}{ll}
      P_i(K_j\mid\omega)\ge
      \frac{u_i(\sigma_i',\sigma_j)-u_i(\tau_i,\sigma_j)}{\left(\frac{u_i(\tau)}{1-\lambda_i(\omega)}-(u_i(\sigma_i',\tau_j)
      +u_i(\sigma)\frac{\lambda_i(\omega)}{1-\lambda_i(\omega)})\right)+(u_i(\sigma_i',\sigma_j)-u_i(\tau_i,\sigma_j))}.
\end{array}
      $$
      or $P_i(K_j\mid\omega)\ge f_i(\omega)$ where $f_i(\omega)$ is defined by:
     $$
\begin{array}{ll}
      f_i(\omega):=
      \max_{\sigma_i'\in F_i}
      \frac{u_i(\sigma_i',\sigma_j)-u_i(\tau_i,\sigma_j)}{\left(\frac{u_i(\tau)}{1-\lambda_i(\omega)}-(u_i(\sigma_i',\tau_j)
      +u_i(\sigma)\frac{\lambda_i(\omega)}{1-\lambda_i(\omega)})\right)+(u_i(\sigma_i',\sigma_j)-u_i(\tau_i,\sigma_j))},
\end{array}
     $$
     where $F_i=\{\sigma_i'\mid u_i(\tau_i,\sigma_j)<u_i(\sigma_i',\sigma_j)\}$.
\end{itemize}
\textbf{Case 2:} $\omega\notin K_i$.

\noindent Player $i$'s payoff under the strategy profile $\eta^*(K_1,K_2)$  is
\begin{eqnarray}
\gamma_i(\eta^*(K_1,K_2)\mid\omega)&=&(1-P_i(K_j\mid\omega))\frac{u_i(\sigma)}{1-\lambda_i(\omega)}+
\nonumber\\
&+&P_i(K_j\mid\omega)\left(u_i(\sigma_i,\tau_j)+u_i(\sigma)\frac{\lambda_i(\omega)}{1-\lambda_i(\omega)}\right).
\nonumber
\end{eqnarray}
\begin{itemize}
  \item Player $i$ can deviate to the strategy $\sigma_i'^*$ described above
        ($\sigma_i'$ may be in the support of $\sigma$. See remark \ref{trivialityremark}(1)).
        His payoff in this case is:
\begin{eqnarray}
        \gamma_i(\sigma_i'^*,\eta_j^*(K_j)\mid\omega)&=&P_i(K_j\mid\omega)\left(u_i(\sigma_i',\tau_j)+
        u_i(\sigma)\frac{\lambda_i(\omega)}{1-\lambda_i(\omega)}\right)
\nonumber\\
        &+&(1-P_i(K_j\mid\omega))\left(u_i(\sigma_i',\sigma_j)+u_i(\sigma)\frac{\lambda_i(\omega)}{1-\lambda_i(\omega)}\right).
\nonumber
\end{eqnarray}
        So that $\eta^*(K_1,K_2)$ is a Bayesian equilibrium, we should have

\noindent $\gamma_i(\eta^*(K_1,K_2)\mid\omega)\ge\gamma_i(\sigma_i'^*,\eta_j^*(K_j)\mid\omega)$, that is
\begin{eqnarray}
\label{eq2}
        P_i(K_j\mid\omega)\big((u_i(\sigma_i,\tau_j)-u_i(\sigma_i',\tau_j))-(u_i(\sigma)-u_i(\sigma_i',\sigma_j))
        \big)+
\nonumber\\
+(u_i(\sigma)-u_i(\sigma_i',\sigma_j))\ge0.
\end{eqnarray}
        The last term on the left side of inequality (\ref{eq2}) is non-negative since $\sigma$ is an equilibrium.
        Therefore, if $u_i(\sigma_i,\tau_j)\ge u_i(\sigma_i',\tau_j)$ inequality \ref{eq2} trivially holds. Inequality (\ref{eq2}) is equivalent to $P_i(K_j\mid\omega)\le g_i^1$, where
        $$
        g_i^1:=\min_{\sigma_i'\in H_i^1}
        \frac{u_i(\sigma)-u_i(\sigma_i',\sigma_j)}
        {(u_i(\sigma)-u_i(\sigma_i',\sigma_j))+(u_i(\sigma_i',\tau_j)-u_i(\sigma_i,\tau_j))},
        $$
        and $H_i^1:=\left\{\sigma_i'\ne\sigma_i,\tau_i\mid u_i(\sigma_i,\tau_j)<u_i(\sigma_i',\tau_j)\right\}$.
        Note that $g_i^1$ is independent of $\omega$ and the inequality is independent of $\lambda_i(\omega)$.
  \item Player $i$ can deviate to $\tau_i^*$. In this case his payoff is:
\begin{eqnarray}
        \gamma_i(\tau_i^*,\eta_j^*(K_j)\mid\omega)&=&P_i(K_j\mid\omega)\frac{u_i(\tau)}{1-\lambda_i(\omega)}+
\nonumber\\
        &+&(1-P_i(K_j\mid\omega))\left(u_i(\tau_i,\sigma_j)+u_i(\sigma)\frac{\lambda_i(\omega)}{1-\lambda_i(\omega)}\right).
\nonumber
\end{eqnarray}
        Because $\gamma_i(\eta^*(K_1,K_2)\mid\omega)\ge\gamma_i(\tau_i^*,\eta_j^*(K_j)\mid\omega)$:
\begin{eqnarray}
\label{eq3}
        P_i(K_j\mid\omega)\left(u_i(\sigma_i,\tau_j)+u_i(\sigma)\left(\frac{\lambda_i(\omega)}{1-\lambda_i(\omega)}\right)
        -\left(\frac{u_i(\tau)}{1-\lambda_i(\omega)}\right)\right.-
\nonumber\\
        \left.-(u_i(\sigma)-u_i(\tau_i,\sigma_j))\right)+
        (u_i(\sigma)-u_i(\tau_i,\sigma_j))\ge0.
\end{eqnarray}
        As before, $u_i(\sigma)-u_i(\tau_i,\sigma_j)\ge0$.
        Unless $u_i(\sigma_i,\tau_j)+u_i(\sigma)\left(\frac{\lambda_i(\omega)}{1-\lambda_i(\omega)}\right)
        -\left(\frac{u_i(\tau)}{1-\lambda_i(\omega)}\right)<0$ inequality (\ref{eq3}) trivially holds.
        Therefore inequality (\ref{eq3}) is equivalent to $P_i(K_j\mid\omega)\le g_i^2(\omega)$, where
        $$
        g_i^2(\omega):=
        \frac{u_i(\sigma)-u_i(\tau_i,\sigma_j)}{(u_i(\sigma)-u_i(\tau_i,\sigma_j))+
        \left(\frac{u_i(\tau)}{1-\lambda_i(\omega)}-(u_i(\sigma_i,\tau_j)
        +u_i(\sigma)\frac{\lambda_i(\omega)}{1-\lambda_i(\omega)})\right)}
        $$
        whenever $u_i(\sigma_i,\tau_j)+u_i(\sigma)\left(\frac{\lambda_i(\omega)}{1-\lambda_i(\omega)}\right)
        <\left(\frac{u_i(\tau)}{1-\lambda_i(\omega)}\right)$, and $g_i^2(\omega):=1$ otherwise.
  \item Player $i$ can play the following strategy $\sigma_i'^{k}$:
        \begin{itemize}
          \item Play $\tau_i$ in the first stage.
          \item If $j$ played $\tau_j$ in the first stage, play $\tau_i$ until
        stage $k+1$ and then play a pure $\sigma_i'\ne\tau_i$, and afterwards $\sigma_i$.
          \item If $j$ played $\sigma_j$ in the first stage, player $i$'s best response is $\sigma_i$ from stage $2$ onwards.
        \end{itemize}
        The payoff is:
$$
\begin{array}{l}
        \gamma_i(\sigma_i'^{k},\eta_j^*(K_j)\mid\omega)=
\\
        \;\;\;\;=P_i(K_j\mid\omega)\left(u_i(\tau)\frac{1-(\lambda_i(\omega))^k}{1-\lambda_i(\omega)}+
        u_i(\sigma_i',\tau_j)(\lambda_i(\omega))^k+
        u_i(\sigma)\frac{(\lambda_i(\omega))^{k+1}}{1-\lambda_i(\omega)}\right)+
\\
        \;\;\;\;\;\;\;\;+(1-P_i(K_j\mid\omega))\left(u_i(\tau_i,\sigma_j)+u_i(\sigma)\frac{\lambda_i(\omega)}{1-\lambda_i(\omega)}\right).
\end{array}
$$
        Note that $\gamma_i(\sigma_i'^{k},\eta_j^*\mid\omega)=A(\lambda_i(\omega))^k+B$ for some constants $A$ and $B$, independent of $k$. Therefore, as a function of $k$, $\gamma_i(\sigma_i'^{k},\eta_j^*(K_j)\mid\omega)$ is monotonic. If it is increasing, it is smaller than the payoff in the previous case (playing $\tau_i^*$). If it is decreasing, $\gamma_i(\eta^*(K_1,K_2)\mid\omega)\ge\gamma_i(\sigma_i'^{k},\eta_j^*(K_j)\mid\omega)$ if and only if the inequality holds for $k=1$. Therefore, we only need to consider the case of $k=1$:
\begin{eqnarray}
\label{eq4}
        P_i(K_j\mid\omega)\left(u_i(\tau_i,\sigma_j)-u_i(\sigma)+u_i(\sigma_i,\tau_j)-u_i(\tau)+\right.
\nonumber\\
        \left.+(u_i(\sigma)-u_i(\sigma_i',\tau_j))(\lambda_i(\omega))
        \right)
        +(u_i(\sigma)-u_i(\tau_i,\sigma_j))&\ge&0.
\end{eqnarray}
        If $u_i(\tau)-u_i(\sigma_i,\tau_j)+(u_i(\sigma_i',\tau_j)-(u_i(\sigma)))\lambda_i(\omega)\ge 0$
        inequality (\ref{eq4}) trivially holds. Therefore inequality (\ref{eq4}) is equivalent to $P_i(K_j\mid\omega)\le g_i^3(\omega)$, where
        $$
\begin{array}{l}
        g_i^3(\omega):=
        \min_{\sigma_i'\in H_i^3(\omega)}
        \frac{u_i(\sigma)-u_i(\tau_i,\sigma_j)}
        {(u_i(\sigma)-u_i(\tau_i,\sigma_j))+(u_i(\tau)-u_i(\sigma_i,\tau_j)+
        (u_i(\sigma_i',\tau_j)-(u_i(\sigma)))(\lambda_i(\omega))},
\end{array}
        $$
        and
        $$
        H_i^3(\omega):=\left\{\sigma_i'\ne\tau_i\mid
        u_i(\tau)-u_i(\sigma_i,\tau_j)+(u_i(\sigma_i',\tau_j)-(u_i(\sigma)))\lambda_i(\omega)<0\right\}.
        $$
\end{itemize}
Then $g_i(\omega):=\min\{g_i^1,g_i^2(\omega),g_i^3(\omega)\}$.
Last, we need to prove that $f_i$ and $g_i$ are $i$ measurable. Both of them are rational functions (or segment-wise rational functions) of $\lambda_i$, and therefore, as functions of $\lambda_i$, they are Borel functions. Since $\Sigma_i$ contains $\{\omega\mid\lambda_i(\omega)\in B\}$, for every open set $B\subseteq[0,1)$, $f_i$ and $g_i$, as functions of $\omega$, are $i$-measurable. $\Box$

\begin{remark}
\label{supportremark}
\begin{enumerate}
  \item If $\tau_i$ is in the support of $\sigma_i$, player $i$ cannot lose by playing $\tau_i$ in every $\omega$:
        $\tau_i$ is a best response to $\sigma_j$ in this case, and player $j$ cannot (in this mechanism)
        discern the deviation and punish.
        Therefore, in this case, in order for $\eta^*(K_1,K_2)$ to be an equilibrium, it is necessary that
        $P_i(K_j\mid\omega)=0$ for every $\omega\notin K_i$.
        Also, in this case $K_i$ is not necessarily a subset of $\Lambda_i$ --- see next remark.
  \item If $\tau_i$ is a best response to $\sigma_j$, $K_i$ may not be a subset of $\Lambda_i$ since we cannot conclude from inequality (\ref{eq1}) that $P_i(K_j\mid\omega)>0$, and therefore we only get that $K_i\subseteq\Lambda_i\cup\{\omega\mid P_i(K_j\mid\omega)=0\}$. Moreover, in this case (and this case only) inequality (\ref{eq1}) trivially holds for every $\omega\in\Lambda_i\cup\{\omega\mid P_i(K_j\mid\omega)=0\}$ --- there is no non-trivial condition on $P_i(K_j\mid\omega)$ that needs to hold for $\omega\in K_i$.
Also, in this case $g_i(\omega)$ may be 0 even if $\tau_i$ isn't in the support of $\sigma_i$ because $g_i^3(\omega)$ and $g_i^2(\omega)$ are either $1$ or $0$ (this follows from their definition because $u_i(\sigma)=u_i(\tau_i,\sigma_j)$). See Example \ref{example best response}.
\end{enumerate}

\end{remark}

\begin{remark}
\label{trivialityremark}

\begin{enumerate}
  \item If there is a best response $\sigma_i'$ to $\sigma_j$ which is a better response to $\tau_j$ than $\sigma_i$, then $g_i\equiv0$ (and then we need $P_i(K_j\mid\omega)=0$ for every $\omega\notin K_i$), even if $\tau_i$ is a best response to $\sigma_j$, because then $g_i^1=0$. Note that this condition depends only on the
        structure of $\Gamma$, and not on the information structure. From this we deduce that if $\sigma_i$ is not pure,
        then $g_i\equiv0$ unless $u_i(\sigma_i',\tau_j)=u_i(\sigma_i'',\tau_j)$ for every $\sigma_i',\sigma_i''$
        in the support of $\sigma_i$.
  \item Under the assumption that $u_i(\tau_i,\sigma_i)<u_i(\sigma)$, we have $f_i>0$. Note that $f_i(\omega)\le1$ for $\omega\in\Lambda_i$, and it can be defined as needed outside this set, see the following remark.
  \item Note that the definition of $f_i$ outside $K_i$ and the definition of $g_i$ inside $K_i$ for $i=1,2$ is
        irrelevant for Theorem \ref{conditionprop}.
\end{enumerate}
\end{remark}

In the following example we show that if $\tau_i$ is a best response to $\sigma_j$, $\eta^*(K_1,K_2)$ may be an equilibrium even if $K_1\supset\Lambda_1$ (a strict inclusion). In particular, the assumption that $\tau_i$ is not a best response to $\sigma_j$ is necessary for Theorem \ref{conditionprop}.

\begin{example}
\label{example best response}
Consider the following game:
\newline \begin{picture}(210,80)(-80,0)
\put( 15,8){$C$}
\put( 15,28){$D$}
\put( 80,50){$D$}
\put(160,50){$C$}
\put( 40, 0){\numbercellong{$1,4$}{}}
\put(40,20){\numbercellong{$1,1$}{}}
\put(120,0){\numbercellong{$3,3$}{}}
\put(120,20){\numbercellong{$4,1$}{}}
\end{picture}
\newline
As in Example \ref{prisonex}, $\sigma=(D,D)$, $\tau=(C,C)$, and $\lambda_1^0=\lambda_2^0=1/3$ (the payoff of player $i$ when he plays $C$ and player $j$ plays $D$ is irrelevant for the calculation of $\lambda_i^0$, as long as $(C,C)$ remains a Nash equilibrium). Let $\Omega=\{1/4,3/4\}^2$, with $\lambda(\omega)=(\lambda_1(\omega),\lambda_2(\omega))=\omega$. In any state of the world $\omega$, player $i$ knows only the coordinate $i$ of $\omega$ (his own discount factor). If $\omega_i=1/4$, player $i$ believes that $\omega=(1/4,1/4)$ (i.e., that both discount factors are 1/4). If $\omega_i=3/4$, player $i$ believes that $\omega_j=1/4$ or $\omega_j=3/4$ with equal probability. From $\lambda_1^0=\lambda_2^0=1/3$ we have that $\Lambda_1=\{3/4\}\times \{1/4,3/4\}$ and $\Lambda_2=\{1/4,3/4\}\times \{3/4\}$. We now show that $\eta^*(\Omega,\Lambda_2)$ is a Bayesian equilibrium:
Player 1 plays $\tau_1^*$ in every state of the world, and player 2 plays $\tau_2^*$ if $\omega_2=3/4$, and always defects otherwise. Therefore, if $\omega_1=1/4$, player 1 believes that $\omega_2=1/4$, so that player 2 always plays $D$, and in this case player 1 gains 1 in each stage regardless of his actions. If $\omega_1=3/4$, player 1 believes that in probability $1/2$ player 2 always plays $D$, so the same argument as before holds, and in probability $1/2$ player 2 plays $\tau_2^*$. Because $\omega_1=3/4>\lambda_1^0$, player 1 cannot profit from deviating from $\tau^*$. Therefore, player 1 cannot profit by deviation.
Player 2 knows that player 1 plays $\tau_1$. If $\omega_2=3/4$ he plays $\tau_2$, and since $3/4>\lambda_2^0$ he cannot profit from deviating. If $\omega_2=1/4$, player 2 plays $D$ in every stage. He cannot profit from playing $\tau_2^*$, since $1/4<\lambda_2^0$. Last, as shown in the proof to Theorem \ref{conditionprop}, we need to check that he cannot profit by deviating to $\sigma_2^{1}$, and indeed inequality (\ref{eq4}) holds in this case (for $\sigma_2'=\sigma_2=D$).
Therefore, $\eta^*(\Omega,\Lambda_2)$ is indeed a Bayesian equilibrium.
\end{example}

\section{Constructing the Cooperation Events}
\label{fbeliefsec}

In this section we construct the cooperation events $K_1$ and $K_2$ so they will satisfy the
conditions of Theorem \ref{conditionprop}.

\subsection{Preliminaries --- Belief Operators}
\label{beliefsec}

In this section we study the concepts $f$-belief and common-$f$-belief, as defined in Definition \ref{definition fbelief}.
This subsection is valid for general belief spaces, not only regarding the model of incomplete information regarding the discount factors (see definition 10.1 in Zamir-Maschler-Solan for a formal definition of a general belief space).

\begin{definition}
The \emph{$f$-belief operator} of player $i$ is the operator $B_i^f:\Sigma \rightarrow \Sigma$ that assigns to each event the states of the world at which player $i$ $f$-believes in the event: $B_i^f(A):=\{\omega\in\Omega \mid P_i(A\mid\omega)\ge f_i(\omega)\}$.
\end{definition}
If $f_i$ is a constant function, and $f_i=f_j$ for every $i$ and $j$, then the concept of $f$-belief reduces to the concept of $p$-belief of Mondrer and Samet (1989). If $f_i$ is a constant function $p_i$, but not necessarily $p_i=p_j$ for every $i$ and $j$, then the concept of $f$-belief reduces to the concept of $\textbf{p}$-belief of  Morris and Kajii (1997).
Similar to the analysis of Mondrer and Samet, we prove the following:

\begin{proposition}
\label{prop1}
Let $B_i^f$ be an $f$-belief operator. The following hold:
\begin{enumerate}
  \item If $A,B\in\Sigma$ and $A\subseteq B$, then $B_i^f(A)\subseteq B_i^f(B)$.
  \item If $A\in\Sigma$ then $B_i^f(B_i^f(A))=B_i^f(A)$.
  \item If $(A_n)_{n=1}^\infty$ is a decreasing sequence of events then $B_i^f(\bigcap_{n=1}^\infty A_n)=\bigcap_{n=1}^\infty B_i^f(A_n)$.
  \item If $C\in\Sigma$ is an $i$-measurable event, then $B_i^f(C)=(C\setminus\{\omega\in\Omega\mid f_i(\omega)>1\})\cup\{\omega\in\Omega\mid f_i(\omega)\le0\}$.
  \item If $f_i>0$ or $\{f_i\le0\}\subseteq C$ then
        $B_i^f(A)\cap C=B_i^f(A\cap C)$ for every event $A\in\Sigma$ and $i$-measurable event $C\in\Sigma$.
\end{enumerate}
\end{proposition}

\noindent\textbf{Proof: } The proof of parts 1 and 3 is similar to the proof for Proposition 2 in Mondrer and Samet
(for $p$-belief operators).
To prove part 4, observe that for every $\omega\in C$, $P_i(C\mid\omega)=1$ and for every $\omega\notin C$, $P_i(C\mid\omega)=0$.
Part 2 follows from part 4, because for every $A\in\Sigma$, $B_i^f(A)$ is an $i$-measurable event (since $f_i$ and $P_i(A\mid\cdot)$ are $i$-measurable), which contains $\{\omega\in\Omega\mid f_i(\omega)\le0\}$.
To prove part 5, assume $\{f_i\le0\}\subseteq C$. In this case $B_i^f(C)=C\setminus\{f_i>1\}$. Assume $\omega\in B_i^f(A\cap C)$. From part 1 we have $\omega\in B_i^f(A)$ and $\omega\in B_i^f(C)\subseteq C$. For the opposite direction, assume $\omega\in B_i^f(A)\cap C$. Then $P_i(A\mid\omega)\ge f_i(\omega)$ and $P_i(C\mid\omega)=1$ and therefore $P_i(A\cap C\mid\omega)\ge f_i(\omega)$, that is, $\omega\in B_i^f(A\cap C)$. $\Box$

We say that \emph{$C$ is a common-$f$-belief at a state of the world $\omega$} if in the state of the world $\omega$ each player $f$-believes $C$, $f$-believes that the other players $f$-believe $C$, $f$-believes that they $f$-believe that he $f$-believes $C$, etc. Therefore, $C$ is a common-$f$-belief at $\omega$, then $\omega\in\bigcap_{i\in N}B_i^f(C)$, $\omega\in\bigcap_{k\in N}B_k^f(\bigcap_{i\in N}B_i^f(C))$, and so on. In particular, if we define $D^0(C):=C$, $D^{n+1}(C):=\bigcap_{i\in N} B_i^f(D^n(C))$ for every $n\ge0$, and $D^f(C):=\bigcap_{n\ge 1} D^n(C)$ we get that ``$C$ is a common-$f$-belief is $\omega$'' is equivalent to $\omega\in D^f(C)$. The event $D^f(C)$ is called \emph{$C$ is a common-$f$-belief}.

By Proposition \ref{prop1}, $B_i^f$ is a belief operator as defined by Mondrer and Samet (1989),
and therefore we get:

\begin{proposition}
\label{prop2}
Let $C\in\Sigma$, $\omega\in\Omega$.
The following definition of ``$C$ is an common-$f$-belief at a state of the world $\omega$'' is equivalent to the Definition \ref{definition fbelief}:
$C$ is a common-$f$-belief at $\omega$ if and only if there exist an event $D\in\Sigma$ such that $\omega\in D$, and $D\subseteq B_i^f(C)$ and $D\subseteq B_i^f(D)$ for every $i\in N$.
\end{proposition}

\subsection{Constructing the Cooperation Events}

Now we go back to the incomplete information structure as defined in Section \ref{section model}.

\begin{definition}
For every two events $C_i\in\Sigma$, $i=1,2$, define
$D_i^{1,f}(C_i,C_j):=B_i^f(C_j)\cap C_i$.
For $n>1$, define

\noindent $D_i^{n,f}(C_i,C_j):=B_i^f(D_j^{n-1,f}(C_j,C_i))\cap D_i^{n-1,f}(C_i,C_j)$,
and

\noindent $D_i^f(C_i,C_j):=\bigcap_{n\ge 1}D_i^{n,f}(C_i,C_j)$.
\end{definition}

The definition of $D_i^f(C_i,C_j)$ is similar (though not identical) to the definition of \emph{iterated $p$-belief} of player $i$ in Morris (1999).

As the next Lemma states, $D_1^f(C_1,C_2)$ and $D_2^f(C_2,C_1)$ are the largest subsets of $C_1$ and $C_2$ (respectively), such that the first inequality in Theorem \ref{conditionprop} holds:

\begin{lemma}
\label{lemma1}
\begin{enumerate}
  \item For every $\omega\in D_i^f(C_i,C_j)$, one has $P_i(D_j^f(C_j,C_i)\mid\omega)\ge f_i(\omega)$ for $i=1,2$.
        $D_1^f(C_1,C_2)$ and $D_2^f(C_2,C_1)$ are the largest subsets of $C_1$ and $C_2$ (respectively) such that
        this property holds.\footnote{They are the largest in the following (strong) sense: if $K_1\subseteq C_1$ and
        $K_2\subseteq C_2$ fulfill $\forall\omega\in K_i\:\: P_i(K_j\mid\omega)\ge f_i(\omega)$ for $i=1,2$ then
        $K_i\subseteq D_i^f(C_i,C_j)$.}
  \item If $C_i$ is $i$-measurable so is $D_i^f(C_i,C_j)$, for every $C_j\in\Sigma$.
  \item If $C_i$ is $i$-measurable and $\{f_i\le0\}\subseteq C_i$ for $i=1,2$, then $D_i^{n,f}(C_i,C_j)$ and $D_i^f(C_i,C_j)$ depend only on the intersection of $C_i$ and $C_j$. In this case, $D_1^f(C_1,C_2)\cap D_2^f(C_2,C_1)=D^f(C_1\cap C_2)$, the event containing all state of the world $\omega$ such that $C_1\cap C_2$ is a common-$f$-belief at $\omega$, and $D_i^f(C_i,C_j)=B_i^f(D^f(C_1\cap C_2))$, the event event containing all state of the world $\omega$ such that player $i$ $f$-believes that $C_1\cap C_2$ is a common-$f$-belief.
\end{enumerate}
\end{lemma}

Note that part 3 holds when $f_1>0$ and $f_2>0$, and in particular to $p$-belief for $p\in(0,1)$.

\noindent\textbf{Proof: }
\begin{enumerate}
  \item We first argue that $D_i^f\subseteq B_i^f(D_j^f)$. Indeed, by Proposition \ref{prop1}(3), and since $(D_j^{n,f})_{n=1}^\infty$ is a decreasing sequence of events,
        $$
        B_i^f(D_j^f)=B_i^f\left(\bigcap_{n\ge 1}D_j^{n,f}\right)=\bigcap_{n\ge 1}B_i^f(D_j^{n,f})\supset
        \bigcap_{n\ge 1}B_i^f(D_j^{n,f})\cap D_i^{n,f}=
        $$
        $$
        =\bigcap_{n\ge 1}D_i^{n+1,f}=D_i^f.
        $$
        Here $D_i^f=D_i^f(C_i,C_j)$ and $D_i^{n,f}=D_i^{n,f}(C_i,C_j)$.
        For the maximality property, assume that that $K_1\subseteq C_1$ and
        $K_2\subseteq C_2$ fulfill $P_i(K_j\mid\omega)\ge f_i(\omega)$ for every $\omega\in K_i$ and $i=1,2$. Therefore
        $K_i\subseteq B_i^f(K_j)$ from which follows $D_i^{1,f}(K_i,K_j)=K_i$ and therefore $D_i^f(K_i,K_j)=K_i$.
        Because $K_i\subseteq C_i$ it follows, from Proposition \ref{prop1}(1) that $D_i^f(K_i,K_j)\subseteq D_i^f(C_i,C_j)$.
  \item This follows from the fact that $B_i^f(C)$ is $i$-measurable for every $C\in\Sigma$, and because the
        intersection of countably many $i$-measurable sets is $i$-measurable.
  \item Denote $C:=C_1\cap C_2$. By Proposition \ref{prop1}(5) one has $D_i^1(C_i,C_j):=B_i^f(C)$.
        This proves the first claim. Because $D_i^1(C_i,C_j):=B_i^f(C)$, it can be verified from the definition of that for $n\ge1$,
        $D^n(C)=D_1^n(C_1,C_2)\cap D_2^n(C_2,C_1)$ and therefore $D^f(C)=D_1^f(C_1,C_2)\cap D_2^f(C_2,C_1)$.
        The event ``player $i$ $f$-believes that $C_1\cap C_2$ is a common-$f$-belief'' is the event
        $$
        B_i^f(D^f(C))=B_i^f(D_1^f(C_1,C_2)\cap D_2^f(C_2,C_1))=
        $$
        $$
        =B_i^f(D_j(C_j,C_i))\cap D_i^f(C_i,C_j)=D_i^f(C_i,C_j).
        $$
        The second equality follows from the second part of this lemma and Proposition \ref{prop1}(5),
        and the last one from the first part of this lemma.
\end{enumerate}
$\Box$

Using Lemma \ref{lemma1} and Proposition \ref{conditionprop} we get the following result:
\begin{theorem}
\label{theorem1}
For every two events $C_1$ and $C_2$, such that $C_i\subseteq\Lambda_i$, for $i=1,2$, and $C_i$ is measurable according to the information of player $i$,
the strategy profile $\eta^*(B_1^f(D^f(C_1\cap C_2),B_2^f(D^f(C_1\cap C_2)))$ is a Bayesian equilibrium if and only if
$B_i^f(D^f(C_1\cap C_2))^c=B_i^{1-g}(D^{1-g}(B_i^f(D^f(C_1\cap C_2))^c\cap B_j^f(D^f(C_1\cap C_2))^c))$.
Moreover, $(B_i^f(D^f(C_1\cap C_2)))_{i=1}^2$ are the maximal subsets of $(C_i)_{i=1}^2$ with this property - if there is $i$ such that $B_i^f(D^f(C_1\cap C_2))\subset K_i\subseteq C_i$, then $\eta^*(K_1,K_2)$ is not an equilibrium.

The opposite is also true: if $\eta^*(K_1,K_2)$ is an equilibrium,
then $B_i^f(D^f(K_i\cap K_j))=K_i$ and $B_i^{1-g}(D^{1-g}(K_i^c\cap K_j^c))=K_i^c$.
\end{theorem}

In other words, $\eta^*(K_1,K_2)$ is an equilibrium if and only if each player either
$f$-believes at $\omega$ that $K_1\cap K_2$ is a common-$f$-belief at $\omega$ or
$1-g$-believes at $\omega$ that the event ``$K_1\cap K_2$ is not a common-$f$-belief'' is a common-$(1-g)$-belief at $\omega$, for every state of the world $\omega\in\Omega$.

Also, if $K_i=B_i^f(D^f(C_1\cap C_2))$ ($i=1,2$) for some $K_i\subseteq C_i\subseteq \Lambda_i$, then each player either
$f$-believes at $\omega$ that $C_1\cap C_2$ is a common-$f$-belief at $\omega$ or
$1-g$-believes at $\omega$ that the event ``$C_1\cap C_2$ is not a common-$f$-belief'' is a common-$(1-g)$-belief at $\omega$, for every state of the world $\omega\in\Omega$.

This theorem shows that the argument that infinite number of conditions of the type ``each player needs to ascribe high enough probability that the other player ascribes high enough probability that his own discount factor is higher than...'' are needed, as was discussed in the introduction, holds. It is equivalent to the condition that in $K_i$ player $i$ $f$-believes $K_j$ at $\omega$ for  every $\omega\in K_i$ for \emph{both} $i=1$ and $i=2$. Also, as the theorem states, a similar condition is required for every $\omega\notin K_i$.

\noindent\textbf{Proof: }
Denote $C=C_1\cap C_2$. From Remark \ref{trivialityremark}(3), we can assure that $\{f_i\le0\}\subseteq D_i^f(C_i,C_j)$ and
$\{1-g_i\le0\}\subseteq D_i^f(C_i,C_j)^c$. Therefore, Lemma \ref{lemma1}(3) holds, so $D_i^f(C_i,C_j)=B_i^f(D^f(C))$ and $D_i^{1-g}(D_i^f(C_i,C_j)^c,D_j^f(C_j,C_i)^c)=B_i^{1-g}(D^{1-g}(B_i^f(D^f(C))^c\cap B_j^f(D^f(C))^c))$.
Because $D_i^f(C_i,C_j)=B_i^f(D^f(C))$, we know from Lemma \ref{lemma1}(1) that the first condition in Proposition \ref{conditionprop} holds,
and that there are no larger subsets of $C_1,C_2$ such that it holds. Therefore, $\eta^*(B_1^f(D^f(C),B_2^f(D^f(C)))$ is an equilibrium if and only if the second condition holds. That is, for every $\omega\notin B_i^f(D^f(C))$ one has $P_i(B_j^f(D^f(C))\mid\omega)\le g_i(\omega)$ which is
equivalent to $P_i(B_j^f(D^f(C))^c\mid\omega)\ge 1-g_i(\omega)$ for every $\omega\in B_i^f(D^f(C))^c$ or
$D_i^f(C_i,C_j)^c=B_i^f(D^f(C))^c\subseteq B_i^{1-g}(D_j^f(C_j,C_i)^c)=B_i^{1-g}(B_j^f(D^f(C))^c)$. From that we get
$$B_i^f(D^f(C))^c=D_i^f(C_i,C_j)^c=D_i^{1,1-g}(D_i^f(C_i,C_j)^c,D_j^f(C_j,C_i)^c),$$
for $i=1,2$, which implies that
$$D_i^f(C_i,C_j)^c=D_i^{n,1-g}(D_i^f(C_i,C_j)^c,D_j^f(C_j,C_i)^c),$$
so that
$$D_i^f(C_i,C_j)^c=D_i^{1-g}(D_i^f(C_i,C_j)^c,D_j^f(C_j,C_i)^c)=$$ $$=B_i^{1-g}(D^{1-g}(B_i^f(D^f(C))^c\cap B_j^f(D^f(C))^c)).$$

For the second part, observe that $B_i^f(D^f(K_i\cap K_j))=K_i$ follows from Lemma \ref{lemma1}(1): $B_i^f(D^f(K_i\cap K_j))=D_i^f(K_i,K_j)$ is the biggest subset of $K_i$ such the first condition in Proposition \ref{conditionprop} holds. But since $\eta^*(K_1,K_2)$ is an equilibrium it holds for $K_i$.
Similarly, $B_i^{1-g}(D^{1-g}(K_i^c\cap K_j^c))=D_i^{1-g}(K_i^c,K_j^c)$ is the biggest subset of $K_i^c$ such that the second condition in Proposition \ref{conditionprop} holds, and therefore $B_i^{1-g}(D^{1-g}(K_i^c\cap K_j^c))=K_i^c$.
$\Box$

\begin{example}
Recall that $\Lambda_i=\{\lambda_i(\cdot)\ge\lambda_i^0\}$. So, if we take $C_i=\Lambda_i$, then
$\eta^*(\Lambda_1,\Lambda_2)$ is an equilibrium if and only if whenever a player does not $f$-believe that "$\lambda_1$ and $\lambda_2$ are high enough" is a common-$f$-belief, he $(1-g)$-believes that the fact that this is
not a common-$f$-belief is a common-$(1-g)$-belief.
\end{example}

\section{Prisoner's Dilemma}
\label{section prisoner}
We now apply the results from previous chapters to the Prisoner's Dilemma,
thus expanding Example \ref{prisonex}.

First we will observe the following regarding a larger class of games:
\begin{lemma}
\label{2actionlemma}
Assume player $i$ has only two actions $\sigma_i$ and $\tau_i$. Then $g_i^1=1$, for $\omega\notin \Lambda_i$, $g_i^2(\omega)=g_i^3(\omega)=1$, and for $\omega\in \Lambda_i$, $g_i^2(\omega)=f_i(\omega)\le g_i^3(\omega)$.
\end{lemma}

\noindent\textbf{Proof: }
$H_i^1=\emptyset$, and therefore $g_i^1=1$. Because there are only two actions, $\Lambda_i=\left\{\omega\in\Omega\mid\: \frac{u_i(\tau)}{1-\lambda_i(\omega)}-(u_i(\sigma_i,\tau_j)+u_i(\sigma)\frac{\lambda_i(\omega)}{1-\lambda_i(\omega)})\ge0\right\}$,
and therefore if $\omega\notin \Lambda_i$ then from the definition of $g_i^2$ it follows that $g_i^2(\omega)=1$.
If $\omega\in \Lambda_i$, it follows that $g_i^2(\omega)=f_i(\omega)$.
In this case the inequality $g_i^2(\omega)\le g_i^3(\omega)$ can be checked arithmetically,
but one can observe that if $\omega\in \Lambda_i$ then player $i$ does not profit by deviating when the profile $\tau^*$ is played,
therefore the deviation that leads to $g_i^2$ is more profitable for him then the deviation that leads to $g_i^3$.

Because $\Lambda_i=\left\{\omega\in\Omega\mid\:
\frac{u_i(\tau)}{1-\lambda_i(\omega)}-(u_i(\sigma_i,\tau_j)+u_i(\sigma)\frac{\lambda_i(\omega)}{1-\lambda_i(\omega)})\ge0\right\}$, if $\omega\notin \Lambda_i$ we have $\frac{u_i(\tau)}{1-\lambda_i(\omega)}-(u_i(\sigma_i,\tau_j)+u_i(\sigma)\frac{\lambda_i(\omega)}{1-\lambda_i(\omega)})<0$
or $u_i(\tau)-u_i(\sigma_i,\tau_j)(1-\lambda_i(\omega))-u_i(\sigma)\lambda_i(\omega)<0$, which implies that $H_i^3=\emptyset$.
$\Box$

\begin{corollary}
\label{2actioncor}
\begin{enumerate}
  \item Let $C_i\subseteq\Lambda_i$ for $i=1,2$. If both players have only two actions, then to verify that $\eta^*(B_1^f(D^f(C_1\cap C_2)),B_2^f(D^f(C_1\cap C_2)))$ is a Bayesian equilibrium, one only needs to verify that $P_i(B_j^f(D^f(C_1\cap C_2))\mid\omega)\le f_i(\omega)$ for every $\omega\in \Lambda_i\setminus B_1^f(D^f(C_1\cap C_2))$ for $i=1,2$.
  \item $\eta^*(B_1^f(D^f(\Lambda)),B_2^f(D^f(\Lambda)))$ is an equilibrium.
\end{enumerate}
\end{corollary}
\noindent Note that in the last case both players cooperate whenever the fact that both players have high enough
discount factor is a common-$f$-belief.

\noindent\textbf{Proof: }
\begin{enumerate}
  \item This follows from Theorem \ref{conditionprop}, Lemma \ref{lemma1}(1) and Lemma \ref{2actionlemma}.
  \item Assume there exists $\omega^*\in \Lambda_1\setminus B_1^f(D^f(\Lambda))$ such that $P_1(B_2^f(D^f(\Lambda)\mid\omega^*)> f_1(\omega^*)$. Denote $K_1:=B_1^f(D^f(\Lambda)\cup\{\omega^*\}$ and $K_2:=B_2^f(D^f(\Lambda)$. From the assumption, $P_i(K_j\mid\omega)\ge f_i(\omega)$ for every $\omega\in K_i$ for $i=1,2$, in contradiction to the maximality property of $D_i^f(\Lambda_i,\Lambda_j)=B_i^f(D^f(\Lambda)$ as shown in Lemma \ref{lemma1}(1). $\Box$
\end{enumerate}

\begin{remark}
If we define $f_i$ to be at least 1 outside $\Lambda_i$ (see Remark \ref{trivialityremark}(3)), Corrolary \ref{2actioncor} holds  for every game which satisfies $f_i\ge g_i$ for $i=1,2$.
\end{remark}

It is easy to calculate that for Prisoner's Dilemma with the payoffs as in Example \ref{prisonex},
$f_i(\omega)=\frac{1-\lambda_i(\omega)}{2\lambda_i(\omega)}$.
Therefore, from the last corollary we have:
\begin{corollary}
\label{prisonercor}
Let $G$ be the repeated Prisoner's Dilemma where each player has incomplete information regarding the other player's
discount factor. Define $f_i(\omega)=\frac{1-\lambda_i(\omega)}{2\lambda_i(\omega)}$.
Then the pair of strategies where each player plays as follows:
If you $f$-believe that it is a common-$f$-belief that
$\lambda_1,\lambda_2\ge\frac{1}{3}$, play the grim-trigger course of action. Otherwise, always defect, is an equilibrium.
\end{corollary}
Therefore, both players will cooperate when $\lambda_1,\lambda_2\ge\frac{1}{3}$ is a common-$f$-belief.

Note that there cannot be larger events than $B_1^f(D^f(\Lambda)),B_2^f(D^f(\Lambda))$ such that the strategy profile $\eta^*$ is an equilibrium, but it may be an equilibrium with smaller ones. See next section for examples.

\section{Examples}

In all the examples in this section, $\Omega\subseteq[0,1)^2$, and we interpret the coordinates of $\omega\in\Omega$ as the players' discount factors, i.e. $\lambda(\omega)=(\lambda_1(\omega),\lambda_2(\omega))=\omega$. In all the examples each player knows his own discount factor (i.e., player $i$ knows coordinate $i$ of $\omega$), and their belief regarding the other player's discount factor depends only on their own discount factor.

In examples \ref{prisonerex1}-\ref{prisonerex3}, $G$ is the repeated Prisoner's Dilemma, with different information structures. In these examples corollaries \ref{2actioncor} and \ref{prisonercor} hold.

In the following example $B_i^f(D^f(\Lambda))\ne\emptyset$, and therefore we get the equilibrium of Corollary \ref{prisonercor}, but we can also get an equilibrium with smaller sets.

\begin{example}
\label{prisonerex1}
Let $\Omega=\{1/4,1/2,3/4\}^2$, and each player assumes a uniform distribution on the other player's discount factor
(regardless of his own discount factor). Equivalently, there is a common-prior of uniform distribution on $\Omega$.

Because $\lambda_1^0=\lambda_2^0=1/3$, it follows that $\Lambda_1=\{1/2,3/4\}\times\{1/4,1/2,3/4\}$ and $\Lambda_2=\{1/4,1/2,3/4\}\times\{1/2,3/4\}$.
Therefore we get that for every $\omega$, $P_1(\Lambda_2\mid\omega)=2/3$. Since $f_1(3/4)<f_1(1/2)=\frac{1-1/2}{2\cdot1/2}=1/2<2/3$
we get $D_1^{1,f}(\Lambda_1,\Lambda_2)=\Lambda_1$. Similarly, $D_2^{1,f}(\Lambda_2,\Lambda_1)=\Lambda_2$,
and therefore $B_1^f(D^f(\Lambda))=D_1^f(\Lambda_1,\Lambda_2)=\Lambda_1$. From Corollary \ref{prisonercor} we get that
$$
\eta_i^*(\omega)=
\left\{\begin{array}{ll}
\tau_i^*&\lambda_i=1/2,3/4,\\
\sigma_i&\lambda_i=1/4,
\end{array}
\right.
$$
defines a Bayesian equilibrium.

Set $C_1=\{3/4\}\times\{1/4,1/2,3/4\}$ and $C_2=\{1/4,1/2,3/4\}\times\{3/4\}$, then $P_1(C_2\mid\omega)=1/3$ for every $\omega$.
Since $f_1(3/4)=1/6>1/3$, we get $D_1^{1,f}(C_1,C_2)=C_1$. Similarly, $D_2^{1,f}(C_2,C_1)=C_2$ ,and therefore $B_i^f(D^f(C_1\cap C_2))=D_i^f(C_i,C_j)=C_i$.
To use Corollary \ref{2actionlemma}(1) we need to verify that
$P_i(C_j\mid\omega)\le f_i(\omega)$ for every $\omega\in \Lambda_i\setminus C_i$. If $\omega\in \Lambda_i\setminus C_i$
then $\lambda_i(\omega)=1/2$, and indeed in $P_i(C_j\mid\omega)=1/3<1/2=f_i(1/2)$. Therefore,
$$
\eta_i^*(\omega)=
\left\{\begin{array}{ll}
\tau_i^*&\lambda_i=3/4,\\
\sigma_i&\lambda_i=1/4,1/2,
\end{array}
\right.
$$
also defines a Bayesian equilibrium.
\end{example}

In the following example $B_i^f(D^f(\Lambda))\ne\emptyset$, and therefore we get the equilibrium of Corollary \ref{prisonercor}, but there are no smaller non-empty sets that define a Bayesian equilibrium.

\begin{example}
\label{prisonerex2}

Let $\Omega$ and $\lambda$ be as in the Example \ref{prisonerex1}, but change the beliefs as follows for $i=1,2$: If $\lambda_i=3/4$, player $i$ believes that $\lambda_j=3/4$, if $\lambda_i=1/2$ he believes that $\lambda_j=1/2$ with probability $1/3$ and that $\lambda_j=3/4$ with probability $2/3$, and if $\lambda_i=1/4$ he believes that $\lambda_j$ is uniform distributed over  $\{1/4,1/2,3/4\}$. Thus, each player believes that the other player has a discount factor at least as high as his own.

$\Lambda_i$ is the same as in Example \ref{prisonerex1}, and $B_i^f(D^f(\Lambda))=\Lambda_i$ so that
$$
\eta_i^*(\omega)=
\left\{\begin{array}{ll}
\tau_i^*&\lambda_i=1/2,3/4,\\
\sigma_i&\lambda_i=1/4,
\end{array}
\right.
$$
defines a Bayesian equilibrium.

If we take $C_i$ as in Example \ref{prisonerex1} we still get $B_i^f(D^f(C_1\cap C_2))=C_i$, but now $P_i(C_j\mid\omega)=2/3>1/2=f_i(\omega)$ for $\omega\in \Lambda_i\setminus C_i$. Therefore
$$
\eta_i^*(\omega)=
\left\{\begin{array}{ll}
\tau_i^*&\lambda_i=3/4,\\
\sigma_i&\lambda_i=1/4,1/2,
\end{array}
\right.
$$
\textbf{does not} define a Bayesian equilibrium. Similarly, $\eta^*(K_1,K_2)$ is not a Bayesian equilibrium for any non-empty $i$-measurable events $K_i\subset\Lambda_i$ (a strict subset), $i=1,2$.
\end{example}

In the following example $B_i^f(D^f(\Lambda))=\emptyset$, and therefore there is no Bayesian equilibrium with
cooperation of the type described here.

\begin{example}
\label{prisonerex0}

Let $\Omega=\{1/4,2/5\}^2$, and $\lambda(\omega)=(\lambda_1(\omega),\lambda_2(\omega))=\omega$. Each player knows his own
discount factor, and if $\lambda_i=2/5$ player $i$ believes that $\lambda_j=1/4$ with probability higher than $1-f_i(2/5)=1/4$ (the rest of the beliefs are irrelevant). Here $\Lambda_i=\{\lambda_i(\cdot)=2/5\}$, and so
$P_i(\Lambda_j\mid\omega)<1/4=f_i(\omega)$ for $\omega\in\Lambda_i$. Therefore $D_i^{1,f}(\Lambda_i,\Lambda_j)=\emptyset$ and
$B_i^f(D^f(\Lambda))=D_i^f(\Lambda_i,\Lambda_j)=\emptyset$, so that we do not get an equilibrium with cooperation of the type $\eta^*$ (because $D_i^f(\Lambda_i,\Lambda_j)$ is the largest set where such an equilibrium is possible, and here it is empty,
we cannot get it for any set). Note that according to Corollary \ref{2actioncor} $\eta^*(B_1^f(D^f(\Lambda)),B_2^f(D^f(\Lambda)))$ is an equilibrium, but in this case, because $D^f(\Lambda)=\emptyset$, it's the trivial equilibrium in which the players always defect in every state of the world $\omega$.
\end{example}

In the following example $B_i^f(D^f(\Lambda))\ne\emptyset$ but $B_i^f(D^f(\Lambda))\ne\Lambda_i$, as was in the previous examples whenever $B_i^f(D^f(\Lambda))$ was non-empty.

\begin{example}
\label{prisonerex3}

Let $\Omega=(0,1)^2$, and assume that each player believes that the other player's discount factor is uniformly distributed, regardless of his own discount factor. $\Sigma$ is the Borel $\sigma$-algebra over $(0,1)^2$. $\Sigma_1$ contains the sets $B\times(0,1)$, for every Borel set $B\subseteq (0,1)$, and $\Sigma_2$ contains the sets $(0,1)\times B$, for every Borel set $B\subseteq (0,1)$.

Here $\Lambda_1=[1/3,1)\times(0,1)$ and $\Lambda_2=(0,1)\times[1/3,1)$. We will see that $B_1^f(D^f(\Lambda))=[1/2,1)\times(0,1)$,
i.e., a strict subset of $\Lambda_1$ (and an analog statement holds for player 2): because $P_1(\Lambda_2\mid\omega)=2/3$ for every $\omega$, and $f_1(3/7)=2/3$ we deduce that $D_1^{1,f}(\Lambda_1,\Lambda_2)=[3/7,1)\times(0,1)$.
Therefore $P_1(D_2^{1,f}\mid\omega)=4/7$ for every $\omega$, and because $f_1(7/15)=4/7$ we deduce that
$D_1^{2,f}(\Lambda_1,\Lambda_2)=[7/15,1)\times(0,1)$. We continue the same way and deduce that for every $k$ there is $n$ such that
$D_1^{k,f}(\Lambda_1,\Lambda_2)=\left[\frac{n}{2n+1},1\right)\times(0,1)$:
$P_1(D_2^k\mid\omega)=1-\frac{n}{2n+1}$ for every $\omega$, and because $f_1\left(\frac{2n+1}{2(2n+1)+1}\right)=1-\frac{n}{2n+1}$
we deduce that $D_1^{k+1,f}(\Lambda_1,\Lambda_2)=\left[\frac{2n+1}{2(2n+1)+1},1\right)\times(0,1)$.
Therefore $B_1^f(D^f(\Lambda))=D_1^f(\Lambda_1,\Lambda_2)=\bigcap_{n\ge 1}D_1^{k,f}=[1/2,1)\times(0,1)$. One can easily check that indeed $P_i(B_j^f(D^f(\Lambda))\mid\omega)\ge f_i(\omega)$ for every $\omega\in B_i^f(D^f(\Lambda))$ for $i=1,2$. Note that here the construction of $D_i^f$ requires an infinite number of steps - for every $k$ we have $D_i^{k+1,f}\ne D_i^{k,f}$.

From Corollary \ref{prisonercor} we deduce that
$$
\eta_i^*(\omega)=
\left\{\begin{array}{ll}
\tau_i^*&\lambda_i\ge1/2,\\
\sigma_i&\lambda_i<1/2,
\end{array}
\right.
$$
defines a Bayesian equilibrium.
\end{example}

\noindent In examples \ref{example5}-\ref{example6}, $G$ is the following repeated game:
\newline \begin{picture}(290,90)(-80,-20)
\put( 15,8){$C$}
\put( 15,28){$D$}
\put( 80,50){$D$}
\put(160,50){$C$}
\put(15,-12){$N$}
\put(240,50){$N$}
\put(40,-20){\numbercellong{$0,\cdot$}{}}
\put( 40, 0){\numbercellong{$0,4$}{}}
\put(40,20){\numbercellong{$1,1$}{}}
\put(120,-20){\numbercellong{$a,\cdot$}{}}
\put(120,0){\numbercellong{$3,3$}{}}
\put(120,20){\numbercellong{$4,0$}{}}
\put(200,-20){\numbercellong{$\cdot,\cdot$}{}}
\put(200,0){\numbercellong{$\cdot,a$}{}}
\put(200,20){\numbercellong{$\cdot,0$}{}}
\end{picture}
\newline
with $a>4$ (payoffs not indicated can be arbitrary).
Here $\sigma=(D,D)$, $\tau=(C,C)$, and $\lambda_1^0=\lambda_2^0=\frac{a-3}{a-1}$.
In these examples Theorem \ref{theorem1} holds, with $f_i(\omega)=g_i^2(\omega)=\frac{1-\lambda_i(\omega)}{2\lambda_i(\omega)}$,
$g_i^1=\frac{1}{a-3}$ and $g_i^3(\omega)=\frac{1}{(a-1)\lambda_i(\omega)}$.

In the following example $B_i^f(D^f(\Lambda))\ne\emptyset$ for $i=1,2$ but these sets does not satisfy the conditions in Theorem \ref{theorem1} (or equivalently those in Theorem \ref{conditionprop}). Moreover, we prove that in this example there are no non-empty $K_1,K_2$ such that $\eta^*(K_1,K_2)$ is a Bayesian equilibrium.

\begin{example}
\label{example5}

Let $a=6$, the belief space as in Example \ref{prisonerex3}. Here $\Lambda_1=[3/5,1)\times(0,1)$ and $\Lambda_2=(0,1)\times[3/5,1)$, and because $f_i(\omega)\le f_i(3/5)=1/3<1-3/5=P_i(\Lambda_j\mid\omega)$ for every $\omega\in\Lambda_i$, $D_i^1(\Lambda_i,\Lambda_j)=\Lambda_i$ and therefore $D_i^f(\Lambda_i,\Lambda_j)=\Lambda_i$.
But, for $\omega\notin\Lambda_i$, $P_i(\lambda_j\mid\omega)=2/5>1/3=g_i^1\ge g_i(\omega)$ and therefore the second condition of Proposition \ref{conditionprop} does not hold, and $\eta^*(B_1^f(D^f(\Lambda),B_2^f(D^f(\Lambda))))$
does not define a Bayesian equilibrium, unlike previous examples.

Now we prove that there are no non-empty $K_1,K_2$ such that $\eta^*(K_1,K_2)$ is a Bayesian equilibrium. Suppose that $\eta^*(K_1,K_2)$ is a Bayesian equilibrium, and that $K_1,K_2$ are non-empty cooperation events. Denote, for $i=1,2$, $\lambda_i^*:=\inf\{\lambda_i(\omega)\mid\omega\in K_i\}$. Since $K_i\subseteq\Lambda_i$, $\lambda_i^*\ge3/5$. Note that since the beliefs are derived from the uniform distribution, $P_i(K_j\mid\omega)$ is independent of the state of the world $\omega$, so we denote it by $P_i(K_j)$. From the first inequality of Theorem \ref{conditionprop}, we have $P_i(K_j)\ge f_i(\omega)$ for every $\omega\in K_i$, and since $f_i$ is continuous, $P_i(K_j)\ge f_i(\lambda_i^*)$. We now argue that if $\lambda_i(\omega)>\lambda_i^*$, then $\omega\in K_i$. Otherwise we have, from the second inequality of Theorem \ref{conditionprop}, that $P_i(K_j)\le g_i(\omega)\le g_i^2(\omega)=f_i(\omega)$. But $f_i(\omega)<f_i(\lambda_i^*)\le P_i(K_j)$, since $\lambda_i(\omega)>\lambda_i^*$, in contradiction. Therefore we have that $P_i(K_j)=1-\lambda_j^*$. Next, we argue that $P_i(K_j)=f_i(\lambda_i^*)$. Otherwise $P_i(K_j)>f_i(\lambda_i^*)$, and there is a state of the world $\omega\in\Omega$ such that
$P_i(K_j)>f_i(\omega)>f_i(\lambda_i^*)$. From the definition of $\lambda_i^*$, we have $\omega\notin K_i$, but then we should have $P_i(K_j)\le g_i^2(\omega)=f_i(\omega)$ in contradiction. We conclude that $1-\lambda_j^*=f_i(\lambda_i^*)$ for $i=1,2$. Because $f_i(\lambda_i)=\frac{1-\lambda_i}{2\lambda_i}$, we have that $\frac{1-\lambda_2^*}{2\lambda_2^*}=1-\lambda_1^*=2\lambda_1^*(1-\lambda_2^*)$, or equivalently, $\lambda_1^*\lambda_2^*=1/4$, in contradiction to $\lambda_i^*\ge3/5$ for $i=1,2$.
\end{example}

In the following example $B_i^f(D^f(\Lambda))\ne\emptyset$ and does not satisfy the condition in Theorem \ref{theorem1},
but for a smaller non-empty set, the conditions hold, and therefore define a Bayesian equilibrium.

\begin{example}
\label{example6}

Let $a=5$, and $\Omega=\{1/4,1/2,3/4\}^2$ with a common-prior of uniform distribution. Here $\lambda_i^0=1/2$ and therefore $\Lambda_i=\{\lambda_i(\cdot)\ge1/2\}$. As in the Example \ref{example5}, $B_i^f(D^f(\Lambda))=D_i^f(\Lambda_i,\Lambda_j)=\Lambda_i$ but
the second condition of Proposition \ref{conditionprop} does not hold because of $g_i^1$:
$P_1(\Lambda_2\mid(1/4,\lambda_2))=2/3>1/2=g_i^1$.

For $C_i=\{\lambda_i=3/4\}$, we get $D_i^f(C_i,C_j)=C_i$ and the second condition of Proposition \ref{conditionprop} does hold (this can easily be verified by calculation). Therefore
$$
\eta_i^*(\omega)=
\left\{\begin{array}{ll}
\tau_i^*& \lambda_i=3/4,\\
\sigma_i& \lambda_i=1/4,1/2,
\end{array}
\right.
$$
defines a Bayesian equilibrium.
\end{example}

\section{$\epsilon$-Equilibria}

In this section, we generalize the results from the previous chapters, for the case of $\epsilon$-equilibria.

First, we prove the following conditions for Bayesian $\epsilon$-equilibrium with cooperation, similar to Theorem \ref{conditionprop}:

\begin{theorem}
\label{epsilonconditionprop}
Let $\epsilon>0$. In the game $G$, the strategy profile $\eta^*(K_1,K_2)=(\eta_1^*(K_1),\eta_2^*(K_2))$, such that $K_1\subseteq\Lambda_1$ and $K_2\subseteq\Lambda_2$, is an $\epsilon$-equilibrium for every $\omega\in\Omega$ if and only if, for $i=1,2$,
\begin{enumerate}
  \item $P_i(K_j\mid\omega)\ge f_i^\epsilon(\omega)$ for every $\omega\in K_i$,
  \item $P_i(K_j\mid\omega)\le g_i^\epsilon(\omega)$ for every $\omega\notin K_i$.
\end{enumerate}
$f_i^\epsilon$ and $g_i^\epsilon$ are $i$-measurable functions that will be described in the proof, and they tend to $f_i$ and $g_i$ when $\epsilon$ tends to 0.
\end{theorem}

\noindent\textbf{Proof: }
The proof is similar to the proof of Theorem \ref{conditionprop}, with the same options for deviation in each case, and the payoffs are the same, only here we assume $\gamma_i(\eta^*(K_1,K_2)\mid\omega)\ge\gamma_i(\cdot,\eta_j^*(K_j)\mid\omega)-\epsilon$ instead of $\gamma_i(\eta^*(K_1,K_2)\mid\omega)\ge\gamma_i(\cdot,\eta_j^*(K_j)\mid\omega)$.
\newline
\textbf{Case 1:} $\omega\in K_i$.
\begin{itemize}
  \item Deviation to $\sigma_i'^{**}$ --- in this case $\gamma_i(\eta^*(K_1,K_2)\mid\omega)\ge\gamma_i(\sigma_i'^{**},\eta_j^*(K_j)\mid\omega)-\epsilon$ trivially holds since $\gamma_i(\eta^*(K_1,K_2)\mid\omega)\ge\gamma_i(\sigma_i'^{**},\eta_j^*(K_j)\mid\omega)$ for every $\omega\in\Lambda_i$.
  \item Deviation to $\sigma_i'^*$ --- here instead of inequality (\ref{eq1}) we get
\begin{eqnarray}
\label{eq epsilon}
      P_i(K_j\mid\omega)\left(\frac{u_i(\tau)}{1-\lambda_i(\omega)}-\left(u_i(\sigma_i',\tau_j)
      +u_i(\sigma)\frac{\lambda_i(\omega)}{1-\lambda_i(\omega)}\right)\right)+
\nonumber\\
      +(1-P_i(K_j\mid\omega))(u_i(\tau_i,\sigma_j)-u_i(\sigma_i',\sigma_j))&\ge-\epsilon.
\end{eqnarray}
or equivalently $P_i(K_j\mid\omega)\ge f_i^\epsilon(\omega)$ where $f_i^\epsilon(\omega)$ is defined by:
     $$
\begin{array}{ll}
      f_i^\epsilon(\omega):=\max_{\sigma_i'\in F_i}
      \frac{u_i(\sigma_i',\sigma_j)-u_i(\tau_i,\sigma_j)-\epsilon}
      {\left(\frac{u_i(\tau)}{1-\lambda_i(\omega)}-(u_i(\sigma_i',\tau_j)
      +u_i(\sigma)\frac{\lambda_i(\omega)}{1-\lambda_i(\omega)})\right)+(u_i(\sigma_i',\sigma_j)-u_i(\tau_i,\sigma_j))},
\end{array}
     $$
where $F_i=\{\sigma_i'\mid u_i(\tau_i,\sigma_j)<u_i(\sigma_i',\sigma_j)\}$.
\end{itemize}
\textbf{Case 2:} $\omega\notin K_i$.
\begin{itemize}
  \item Deviation to $\sigma_i'^*$ --- here instead of inequality (\ref{eq2}) we get
\begin{eqnarray}
        P_i(K_j\mid\omega)\big((u_i(\sigma_i,\tau_j)-u_i(\sigma_i',\tau_j))-(u_i(\sigma)-u_i(\sigma_i',\sigma_j))
        \big)+
\nonumber\\
        +(u_i(\sigma)-u_i(\sigma_i',\sigma_j)&\ge-\epsilon,
\nonumber
\end{eqnarray}
or equivalently $P_i(K_j\mid\omega)\le g_i^{1\epsilon}$, where
        $$
        g_i^{1\epsilon}:=\min_{\sigma_i'\in H_i^1}
        \frac{u_i(\sigma)-u_i(\sigma_i',\sigma_j)+\epsilon}
        {(u_i(\sigma)-u_i(\sigma_i',\sigma_j))+(u_i(\sigma_i',\tau_j)-u_i(\sigma_i,\tau_j))},
        $$
        and $H_i^1:=\left\{\sigma_i'\ne\sigma_i,\tau_i\mid u_i(\sigma_i,\tau_j)<u_i(\sigma_i',\tau_j)\right\}$.
  \item Deviation to $\tau_i$ --- here instead of inequality (\ref{eq3}) we get
\begin{eqnarray}
        P_i(K_j\mid\omega)\left(u_i(\sigma_i,\tau_j)+u_i(\sigma)\left(\frac{\lambda_i(\omega)}{1-\lambda_i(\omega)}\right)
        -\left(\frac{u_i(\tau)}{1-\lambda_i(\omega)}\right)\right.-
\nonumber\\
        \left.-(u_i(\sigma)-u_i(\tau_i,\sigma_j))\right)+
        +(u_i(\sigma)-u_i(\tau_i,\sigma_j))&\ge-\epsilon,
\nonumber
\end{eqnarray}
or equivalently $P_i(K_j\mid\omega)\le g_i^{2\epsilon}(\omega)$ where
        $$
        g_i^{2\epsilon}(\omega):=
        \frac{u_i(\sigma)-u_i(\tau_i,\sigma_j)+\epsilon}{(u_i(\sigma)-u_i(\tau_i,\sigma_j))+
        \left(\frac{u_i(\tau)}{1-\lambda_i(\omega)}-(u_i(\sigma_i,\tau_j)
        +u_i(\sigma)\frac{\lambda_i(\omega)}{1-\lambda_i(\omega)})\right)}
        $$
        whenever $u_i(\sigma_i,\tau_j)+u_i(\sigma)\left(\frac{\lambda_i(\omega)}{1-\lambda_i(\omega)}\right)
        <\left(\frac{u_i(\tau)}{1-\lambda_i(\omega)}\right)$ and $g_i^{2\epsilon}(\omega):=1$ otherwise.
  \item Deviation to $\sigma_i'^{k}$ --- here instead of inequality (\ref{eq4}) we get
\begin{eqnarray}
        P_i(K_j\mid\omega)\left(u_i(\tau_i,\sigma_j)-u_i(\sigma)+u_i(\sigma_i,\tau_j)-u_i(\tau)+\right.
\nonumber\\
        \left.+(u_i(\sigma)-u_i(\sigma_i',\tau_j))(\lambda_i(\omega))
        \right)
        +(u_i(\sigma)-u_i(\tau_i,\sigma_j))&\ge-\epsilon,
\nonumber
\end{eqnarray}
or equivalently $P_i(K_j\mid\omega)\le g_i^{3\epsilon}(\omega)$ where
        $$
\begin{array}{ll}
        g_i^{3\epsilon}(\omega):=
        \min_{\sigma_i'\in H_i^3(\omega)}
        \frac{u_i(\sigma)-u_i(\tau_i,\sigma_j)+\epsilon}
        {(u_i(\sigma)-u_i(\tau_i,\sigma_j))+(u_i(\tau)-u_i(\sigma_i,\tau_j)+
        (u_i(\sigma_i',\tau_j)-(u_i(\sigma)))(\lambda_i(\omega))}
\end{array}
        $$
        and
        $$
        H_i^3(\omega):=\left\{\sigma_i'\ne\tau_i\mid
        u_i(\tau)-u_i(\sigma_i,\tau_j)+(u_i(\sigma_i',\tau_j)-(u_i(\sigma)))\lambda_i(\omega)<0\right\}.
        $$
\end{itemize}
Set $g_i^\epsilon(\omega):=\min\{g_i^{1\epsilon},g_i^{2\epsilon}(\omega),g_i^{3\epsilon}(\omega)\}$. The proof that $f_i^\epsilon$ and $g_i^\epsilon$ are $i$-measurable is the same as in the proof of Theorem \ref{conditionprop}. $\Box$

\begin{remark}
Note that the conditions given in this theorem are under the assumption that $K_1\subseteq\Lambda_1$ and $K_2\subseteq\Lambda_2$, and therefore the theorem does not describe necessary conditions for $\eta^*(K_1,K_2)$ to be an $\epsilon$-equilibrium. For example, define
$\lambda_i^\epsilon:=\min\left\{\lambda_i\mid
\frac{u_i(\tau)}{1-\lambda_i}-(u_i(\sigma_i',\tau_j)+u_i(\sigma)\frac{\lambda_i}{1-\lambda_i})\ge-\epsilon
\;\;\forall \sigma_i'\ne\tau_i\right\}$ and $\Lambda^\epsilon:=\{\omega\in\Omega\mid\lambda_i(\omega)\ge\lambda_i^\epsilon\;\;i=1,2\}$.
Then, when the information regarding the discount factors is complete, $\eta^*(\Lambda^\epsilon,\Lambda^\epsilon)$ is an $\epsilon$-equilibrium, but generally $\Lambda^\epsilon\not\subseteq\Lambda_i$ for $i=1,2$.
\end{remark}

\begin{corollary}
Theorem \ref{theorem1} holds for $\epsilon$-equilibrium with the following adjustments: First, instead of $f$ and $g$ there should be $f^\epsilon$ and $g^\epsilon$; and second, in the second part of the theorem, the assumption that $K_i\subseteq\Lambda_i$ for $i=1,2$ should be added.
\end{corollary}

\noindent\textbf{Proof: } The proof is exactly like the proof of Theorem \ref{theorem1}. $\Box$

The following lemma shows that to $f^\epsilon$-believe in an event or to $1-g^\epsilon$-believe in an event are weaker demands than to $p$-believe in an event, for $p<1$ sufficiently large. This property will be used in Section \ref{section almost}.

\begin{lemma}
\label{epsilonprop}
Let $K_i\subseteq\Lambda_i$ for $i=1,2$. For $\epsilon>0$ small enough, there exist constants $M_i,N_i>0$ such that $0<f_i^\epsilon(\omega)<1-\epsilon/M_i$ for every $\omega\in K_i$ and $g_i^\epsilon(\omega)>\epsilon/N_i$ for every $\omega\notin K_i$.
\end{lemma}

\noindent\textbf{Proof: }
Let $\sigma_i''$ be player $i$'s best response to $\tau_j$ in $\Gamma$, and $\sigma_i'''$ the worst response. It follows, from the definition of $f_i^\epsilon$, that $$f_i^\epsilon(\omega)<f_i(\omega)-\frac{\epsilon}{\left(\frac{u_i(\tau)}{1-\lambda_i(\omega)}-(u_i(\sigma_i''',\tau_j)
      +u_i(\sigma)\frac{\lambda_i(\omega)}{1-\lambda_i(\omega)})\right)+(u_i(\sigma)-u_i(\tau_i,\sigma_j))}.$$
Denote $M_i:=2(u_i(\sigma)-u_i(\tau_i,\sigma_j))$. There exist $\lambda_i^1>\lambda_i^0$ such that, for every $\lambda_i\in[\lambda_i^0,\lambda_i^1]$, $\frac{u_i(\tau)}{1-\lambda_i(\omega)}-(u_i(\sigma_i''',\tau_j)
      +u_i(\sigma)\frac{\lambda_i(\omega)}{1-\lambda_i(\omega)})<M_i/2$. Therefore, for every $\omega\in K_i$ such that $\lambda_i(\omega)\in[\lambda_i^0,\lambda_i^1]$, $f_i^\epsilon(\omega)<1-\epsilon/M_i$. If $\lambda_i(\omega)>\lambda_i^1$, we get that
\begin{eqnarray}
\frac{u_i(\tau)}{1-\lambda_i(\omega)}-\left(u_i(\sigma_i'',\tau_j)+u_i(\sigma)\frac{\lambda_i(\omega)}{1-\lambda_i(\omega)}\right)=
\nonumber\\
=u_i(\tau)\left(\frac{1}{1-\lambda_i(\omega)}-\frac{1}{1-\lambda_i^0}\right)-
u_i(\sigma)\left(\frac{\lambda_i(\omega)}{1-\lambda_i(\omega)}-\frac{\lambda_i^0}{1-\lambda_i^0}\right)>
\nonumber\\
>(u_i(\tau)-u_i(\sigma))\left(\frac{1}{1-\lambda_i^1}-\frac{1}{1-\lambda_i^0}\right)=:M,
\nonumber
\end{eqnarray}
and therefore $f_i^\epsilon(\omega)<\frac{M_1/2}{M+M_1/2}$. For $\epsilon$ small enough we will get $f_i^\epsilon(\omega)<1-\epsilon/M_i$ for every $\omega\in K_i$.
$f_i^\epsilon(\omega)>0$ for every $\omega\in K_i$ if $\epsilon<u_i(\sigma)-u_i(\tau_i,\sigma_j)$.
From the definitions of $g_i^{1\epsilon}$, $g_i^{2\epsilon}$ and $g_i^{3\epsilon}$ we get that, for every $\omega\notin K_i$,
$$
        g_i^{1\epsilon}>\min_{\sigma_i'\in H_i^1}
        \frac{\epsilon}
        {u_i(\sigma_i',\tau_j)-u_i(\sigma_i,\tau_j)},
        $$
and
$$
        g_i^{2\epsilon}(\omega),g_i^{3\epsilon}(\omega)>
        \frac{\epsilon}{u_i(\sigma)-u_i(\tau_i,\sigma_j)}.
        $$
Therefore $g_i^\epsilon(\omega)>\epsilon/N_i$ for every $\omega\notin K_i$. $\Box$

Next, we generalize the results of Section \ref{section prisoner} to $\epsilon$-equilibria in games where both players have only two actions.

The following lemma is the $\epsilon$-equilibrium equivalent of Lemma \ref{2actionlemma}:
\begin{lemma}
\label{2actionlemma epsilon}
Assume player $i$ has only two actions $\sigma_i$ and $\tau_i$. Then, for every $\epsilon>0$, $g_i^{1,\epsilon}=1$, for $\omega\notin \Lambda_i$, $g_i^{2,\epsilon}(\omega),g_i^3(\omega)\ge1$, and for $\omega\in \Lambda_i$, $g_i^{2,\epsilon}(\omega)>f_i^\epsilon(\omega)$ and $g_i^{2,\epsilon}(\omega)\ge g_i^{3,\epsilon}(\omega)$.
\end{lemma}

\noindent\textbf{Proof: } All the results, except $g_i^{2,\epsilon}(\omega)\ge g_i^{3,\epsilon}(\omega)$ for $\omega\in \Lambda_i$, follow from Lemma \ref{2actionlemma}, since $g_i^{k,\epsilon}\ge g_i^k$ for $k=1,2,3$ and $f_i\ge f_i^\epsilon$. The same reasoning that shows $g_i^{2}(\omega)\ge g_i^{3}(\omega)$  for $\omega\in \Lambda_i$, $g_i^{2,\epsilon}(\omega)\ge g_i^{3,\epsilon}(\omega)$. $\Box$

The following corollary is the $\epsilon$-equilibrium equivalent of Corollary \ref{2actioncor}:
\begin{corollary}
\label{2actioncor epsilon}
\begin{enumerate}
  \item Let $C_i\subseteq\Lambda_i$ for $i=1,2$. If both players have only two actions, then to verify that $\eta^*(B_1^{f^\epsilon}(D^{f^\epsilon}(C_1\cap C_2)),B_2^{f^\epsilon}(D^{f^\epsilon}(C_1\cap C_2))$ is a Bayesian $\epsilon$-equilibrium, it is sufficient to verify that $P_i(B_j^{f^\epsilon}(D^{f^\epsilon}(C_1\cap C_2))\mid\omega)\le f_i^\epsilon(\omega)$ for every $\omega\in \Lambda_i\setminus B_i^{f^\epsilon}(D^{f^\epsilon}(C_1\cap C_2))$ for $i=1,2$.
  \item $\eta^*(B_1^{f^\epsilon}(D^{f^\epsilon}(\Lambda)),B_2^{f^\epsilon}(D^{f^\epsilon}(\Lambda)))$ is an $\epsilon$-equilibrium.
\end{enumerate}
\end{corollary}

\noindent\textbf{Proof: }
\begin{enumerate}
  \item This follows from Theorem \ref{epsilonconditionprop}, Lemma \ref{lemma1}(1) and Lemma \ref{2actionlemma epsilon}.
  \item The proof is similar to the proof of Corollary \ref{2actioncor}(2), only now it follows from Corollary \ref{2actioncor epsilon}(1) instead from Corollary \ref{2actioncor}(1). $\Box$
\end{enumerate}

\section{``Almost'' Complete Information}
\label{section almost}

In this section we look at belief spaces that are ``almost'' complete, in several senses, and see whether the cooperation events when the information is almost complete are close to the cooperation events when the information is complete.

\subsection{Knowing Approximately the Discount Factors}

The first sense of ``almost'' complete information we consider, is when each player knows the other's discount factor, up to a range of $\epsilon>0$. That is, player $i$ knows his own discount factor, and is given a signal $x$ so he knows that player $j$'s discount factor is between $x-\epsilon$ and $x+\epsilon$. As the following example shows, in this case the cooperation events are significantly different from the complete information cooperation events, even in the repeated Prisoner's Dilemma, regardless of $\epsilon$, and even if we consider only $\delta$-equilibrium for some $\delta>0$.

\begin{example}
\label{prisonerex4}
Let $G$ be the repeated Prisoner's Dilemma.
Let $\Omega=(0,1)^2$, $\Sigma$ is the Borel $\sigma$-algebra, and as in previous examples, $\lambda(\omega)=(\lambda_1(\omega),\lambda_2(\omega))=\omega$ for every $\omega\in\Omega$. Each player knows his own discount factor, and believes that the other's discount factor
is within $\epsilon>0$ of his own; if $\lambda_i(\omega)=x$, then player $i$ believes that $\lambda_j$ is uniformly
distributed in $(x-\epsilon,x+\epsilon)$ (with cutoffs if one side exceeds $0$ or $1$).
Again, $\Lambda_1=[1/3,1)\times(0,1)$ and $\Lambda_2=(0,1)\times[1/3,1)$.
We will show that $B_1^f(D^f(\Lambda))=[1/2,1)\times(0,1)$ regardless of $\epsilon$ (for $\epsilon$ small enough):

Let $\delta>0$. Because $f_i$ is a continuous monotonically-decreasing function, and because $f_i(1/2)=(1/2)$, there
exists $\delta'>0$ such that $f_i(1/2-\delta')=1/2+\delta$. Therefore, any state of the world $\omega$ such that
$\lambda_i(\omega)<\min\{1/3+2\epsilon\delta,1/2-\delta'\}$ is not in $D_i^{1,f}(\Lambda_i,\Lambda_j)$. That is because
for such states of the world $\omega$, one has $P_i(\Lambda_j\mid\omega)=P_i(\lambda_j\ge1/3\mid\omega)<1/2+\delta=f_i(1/2-\delta')<f_i(\omega)$.
Therefore, for some $x^1\ge\min\{1/3+2\epsilon\delta,1/2-\delta'\}$, $D_1^{1,f}(\Lambda_1,\Lambda_2)=[x^1,1)\times(0,1)$, and similarly for player 2. Therefore, any state of the world $\omega$ such that $\lambda_i(\omega)<\min\{x^1+2\epsilon\delta,1/2-\delta'\}$ is not in $D_i^{2,f}(\Lambda_i,\Lambda_j)$ by the same reasoning, and $D_1^{2,f}(\Lambda_1,\Lambda_2)=[x^2,1)\times(0,1)$ for $x^2\ge\min\{x^1+2\epsilon\delta,1/2-\delta'\}$.
We continue until we get that any state of the world $\omega$ such that $\lambda_i(\omega)<1/2-\delta'$ is not in $D_i^f(\Lambda_i,\Lambda_j)$.
This is true for every $\delta>0$. Because $\delta'$ tends to $0$ as $\delta$ tends to $0$,
$D_1^f(\Lambda_1,\Lambda_2)\subseteq[1/2,1)\times(0,1)$ (similarly for player 2). The equality follows from Lemma
\ref{lemma1}(1), because $[1/2,1)\times(0,1)$ and $(0,1)\times[1/2,1)$ fulfill the inequality.

We get from Corollary \ref{prisonercor} that
$$
\eta_i^*(\omega)=
\left\{\begin{array}{ll}
\tau_i^*&\lambda_i\ge1/2,\\
\sigma_i&\lambda_i<1/2,
\end{array}
\right.
$$
defines a Bayesian equilibrium and that there is not such an equilibrium with larger sets of
cooperation, regardless of $\epsilon$. Therefore, under this profile the players cooperate when both discount factors are no less than $1/2$, regardless of $\epsilon$, whereas is the complete information case they can cooperate whenever both discount factors are no less than $1/3$ (see Example \ref{prisonex}).

Note that this analysis does not change if $\Omega=(0,1)^2\cap\{(x,y):|x-y|<\epsilon\}$, which verifies that the true state of the world is within the support of the beliefs of the players. Also, if for every state of the world $\omega$ player $i$ believes that $\lambda_j$ is distributed in any non-atomic symmetric way around $\lambda_i(\omega)$, the result still holds.

For a Bayesian $\delta$-equilibrium we follow the same route, with $f^\delta$ instead of $f$ (see Corollary \ref{2actioncor epsilon}(2)). Instead of a threshold of $1/2$, we show that $B_1^{f^\delta}(D^{f^\delta}(\Lambda))=\left[\frac{1-\delta}{2},1\right)\times(0,1)$. That is because $f_i^\delta(\frac{1-\delta}{2})=1-\frac{1-\delta}{2}$. We get a lower threshold, which is still independent of $\epsilon$, and, for low enough $\delta$, higher than $1/3$.
\end{example}

\subsection{The True Discount Factors are Common-$(1-\epsilon)$-Belief in Most States of the World}
\label{section almost complete MS}

In this section we assume a common prior $P$ over the states of the world.
\begin{definition}
\label{defin almost comp MS}
Let $\epsilon,\delta>0$. We say that the discount factors are \emph{almost complete information} with respect to $\epsilon$ and $\delta$, if the set of states of the world in which the true discount factors are common-$(1-\epsilon)$-belief has probability at least $1-\delta$.
\end{definition}

From Theorem B in Mondrer and Samet (1989), we have that if the number of states of the world is finite, then in this case there are a strategy profile $\eta$ and an event $\Omega'$ with probability at least $(1-2\epsilon)(1-\delta)$, such that: (a) $\eta(\omega)=\eta^*(\Lambda,\Lambda)(\omega)$ for every $\omega\in\Omega'$; and (b) $\eta$ is an $\epsilon'$-equilibrium for $\epsilon'>4M\epsilon$ ($M$ is the maximum of the absolute value of the payoffs in the repeated game, taken over all the discount factors). In other words, there is an $\epsilon'$-equilibrium profile that coincides over a large set with the conditional-grim-trigger equilibrium of maximum cooperation in the complete information case.

However, as Example \ref{ex no complete 3} shows, this profile $\eta$ may not be a conditional-grim-trigger profile. It shows that there may be no conditional-grim-trigger $\epsilon'$-equilibria whatsoever, unless there are only two actions for each player (see Corollary \ref{cor 2 action almost}).

\begin{example}
\label{ex no complete 3}
Let $G$ be the game as in Example \ref{example6}.
Let the information structure be as follows: each player has two possible discount factors. That is, $$S=\{(H_1,H_2),(H_1,L_2),(L_1,H_2),(L_1,L_2)\},$$
where $H_i$ is the higher discount factor of player $i$, and $L_i$ the lower one. Assume $L_i<1/3<\lambda_i^0<H_i$ for $i=1,2$. In every state of nature, each player gets a signal regarding the other player's discount factor, which may be either $h_i$ or $l_i$, that is $\Omega=S\times\{h_1,l_1\}\times\{h_2,l_2\}$, the first two coordinates are the discount factors, and the last two are the signals. Each player knows only his discount factor and his signal.
\newline
The common prior on $\Omega$ is as follows: the state of nature is chosen according to the distribution $$\left[(1-\delta)(H_1,H_2),\frac{\delta}{3}(L_1,L_2),\frac{\delta}{3}(L_1,H_2),\frac{\delta}{3}(H_1,L_2)\right].$$
In every state of nature, there is a probability $1-\epsilon$ that both the signals are correct, and the other three states of the world that correspond to this state of nature have equal probability $\epsilon/3$. For example, if $(H_1,H_2)$ is chosen, the distribution over the states of the world is
$$
\left[(1-\epsilon)(H_1,H_2,h_1,h_2),\frac{\epsilon}{3}(H_1,H_2,l_1,h_2),
\frac{\epsilon}{3}(H_1,H_2,h_1,l_2),\frac{\epsilon}{3}(H_1,H_2,l_1,l_2)\right].
$$
Assume that $\delta\ll\epsilon$.
\newline
We now calculate the beliefs of player 1, up to normalization, that is, the sum of the coefficients in the following equations may not be 1. From symmetry, the beliefs of player 2 are similar.
$$
\begin{array}{l}
P_1((H_1,\cdot,h_1,\cdot))=
\\
\;\;\;\;\left[((1-\epsilon)(1-\delta))(H_1,H_2,h_1,h_2),\frac{(1-\delta)\epsilon}{3}(H_1,H_2,h_1,l_2)\right.,
\\
\;\;\;\;\left.\frac{\delta\epsilon}{9}(H_1,L_2,h_1,h_2),\frac{\delta\epsilon}{9}(H_1,L_2,h_1,l_2)\right],
\end{array}
$$
$$
\begin{array}{l}
P_1((L_1,\cdot,l_1,\cdot))=
\\
\;\;\;\;\left[\frac{(1-\epsilon)\delta}{3}(L_1,L_2,l_1,l_2),\frac{\delta\epsilon}{9}(L_1,L_2,l_1,h_2)\right.,
\\
\;\;\;\;\left.\frac{\delta\epsilon}{9}(L_1,H_2,l_1,l_2),\frac{\delta\epsilon}{9}(L_1,H_2,l_1,h_2)\right],
\end{array}
$$
$$
\begin{array}{l}
P_1((H_1,\cdot,l_1,\cdot))=
\\
\;\;\;\;\left[\frac{(1-\delta)\epsilon}{3}(H_1,H_2,l_1,h_2),\frac{(1-\delta)\epsilon}{3}(H_1,H_2,l_1,l_2),\right.
\\
\;\;\;\;\left.\frac{\delta\epsilon}{9}(H_1,L_2,l_1,l_2),\frac{(1-\epsilon)\delta}{3}(H_1,L_2,l_1,h_2)\right],
\end{array}
$$
and
$$
\begin{array}{l}
P_1((L_1,\cdot,h_1,\cdot))=
\\
\;\;\;\;\left[\frac{\delta\epsilon}{9}(L_1,L_2,h_1,l_2),\frac{\delta\epsilon}{9}(L_1,L_2,h_1,h_2),\right.
\\
\;\;\;\;\left.\frac{\delta\epsilon}{9}(L_1,H_2,h_1,h_2),\frac{(1-\epsilon)\delta}{3}(L_1,H_2,h_1,l_2)\right].
\end{array}
$$
Note that for $\omega=(H_1,H_2,h_1,h_2)$, which happens with probability $(1-\epsilon)(1-\delta)$, the true state of nature is a common-$(1-\epsilon)$-belief, and therefore the discount factors are almost complete information with respect to $\epsilon$ and $\epsilon+\delta$, according to Definition \ref{defin almost comp MS}.

We now prove that there is no conditional-grim-trigger $\epsilon'$-equilibrium for small enough $\epsilon'$, independent of $\epsilon$ and $\delta$. Let $\epsilon'$ be such that $g_i^{\epsilon'}(L_i)<1$ and $f_i^{\epsilon'}(H_i)>0$ for $i=1,2$. These inequalities are satisfied by $\epsilon'$ low enough, since (a) $\lim_{\epsilon'\rightarrow 0}f_i^{\epsilon'}=f_i$, $\lim_{\epsilon'\rightarrow 0}g_i^{\epsilon'}=g_i$, and (b) $f_i(H_i)>0$ and $g_i(L_i)<1$ because $L_i<\lambda_i^0<H_i$. Note that since $L_i<\lambda_i^0<H_i$ for $i=1,2$, we have $\Lambda_1=\{(H_1,\cdot,\cdot,\cdot)\}$ and $\Lambda_2=\{(\cdot,H_2,\cdot,\cdot)\}$. Assume also that for $i=1,2$, $\epsilon'<1-2L_i/(1-L_i)$ (we assumed $1/3>L_i$ so $1>2L_i/(1-L_i)$). In this case Inequality (\ref{eq epsilon}) does not hold for $\sigma'=D$, for any $P_i(K_j\mid\omega)$.
Suppose $\eta^*(K_1,K_2)$ is an $\epsilon'$-equilibrium profile with non-trivial cooperation events. Since In this case Inequality (\ref{eq epsilon}) does not hold for player $i$ when his discount factor is $L_i$, we have $K_i\subseteq\Lambda_i$ for $i=1,2$, and so the conditions of Theorem \ref{epsilonconditionprop} holds.
Suppose that $\{(\cdot,H_2,\cdot,l_2)\}\subseteq K_2$. Then $P_1(K_2\mid (L_1,H_2,h_1,l_2))=1-\epsilon>g_1^{\epsilon'}(L_1)$ for $\epsilon$ low enough, which is a contradiction since $(L_1,H_2,h_1,l_2)\notin K_1$. Therefore $K_2=\{(\cdot,H_2,\cdot,h_2)\}$, an by a similar argument $K_1=\{(H_1,\cdot,h_1,\cdot)\}$. But then we have $P_1(K_2|(H_1,H_2,h_1,l_2))<\epsilon<f_1^{\epsilon'}(H_1)$ for low enough $\epsilon$, in contradiction to $\eta^*(K_1,K_2)$ being an $\epsilon'$-equilibrium. Therefore there is no conditional-grim-trigger $\epsilon'$-equilibrium profile with non-trivial cooperation events.
\end{example}

The following lemma is a result of this concept of almost complete information:
\begin{lemma}
\label{lemma almost MS}
Set $\epsilon>0$ and $\delta>0$, and assume that the discount factors are almost complete information with respect to $\epsilon$ and $\delta$. Then $P(\Lambda\setminus D^{1-\epsilon}(\Lambda))<\delta$.
\end{lemma}

\noindent\textbf{Proof: }
Denote $K=\{\omega\in\Omega \mid \lambda(\omega) \mbox{ is not a common-$(1-\epsilon)$-belief in } \omega\}$. From our assumption $P(K)<\delta$. Since $\Lambda\setminus D^{1-\epsilon}(\Lambda)\subseteq K$, we have $P(\Lambda\setminus D^{1-\epsilon}(\Lambda))<\delta$. $\Box$

In the case where each player has two actions, Lemma \ref{lemma almost MS} implies that there is an $\epsilon'$-equilibrium with cooperation events close to $\Lambda$:
\begin{corollary}
\label{cor 2 action almost}
Suppose each player has two actions. Let $\epsilon>0$ and $\delta>0$ and $M:=2\max_{i=1,2}(u_i(\sigma)-u_i(\tau_i,\sigma_j))$. Assume that the discount factors are almost complete information with respect to $\epsilon$ and $\delta$. Then, for every $\epsilon'\ge M\epsilon$, the strategy profile $\eta^*(B_1^{f^{\epsilon'}}(D^{f^{\epsilon'}}(\Lambda)),B_2^{f^{\epsilon'}}(D^{f^{\epsilon'}}(\Lambda)))$ is an $\epsilon'$-equilibrium,
and $P(\Lambda\setminus D^{f^{\epsilon'}}(\Lambda))<\delta$.
\end{corollary}

\noindent\textbf{Proof: }
From Corollary \ref{2actionlemma epsilon}(2) we have that $\eta^*(B_1^{f^{\epsilon'}}(D^{f^{\epsilon'}}(\Lambda)),B_2^{f^{\epsilon'}}(D^{f^{\epsilon'}}(\Lambda)))$ is an $\epsilon'$-equilibrium. From Lemma \ref{epsilonprop} we have that $f_i^{\epsilon'}(\omega)<1-\epsilon$ for every $\omega\in B_i^{f^{\epsilon'}}(D^{f^{\epsilon'}}(\Lambda))$. Therefore $D^{1-\epsilon}(\Lambda)\subseteq D^{f^{\epsilon'}}(\Lambda)\subseteq\Lambda$, so from Lemma \ref{lemma almost MS} we have that $P(\Lambda\setminus D^{f^{\epsilon'}}(\Lambda))<\delta$. $\Box$

\subsection{Each Player $(1-\epsilon)$-Believes that a State of Nature is Common-$(1-\epsilon)$-Belief}

In this section we assume that the information is almost complete in a different sense:
\begin{definition}
\label{defin almost comp strong}
Let $\epsilon>0$. We say that the discount factors are \emph{almost complete information} with respect to $\epsilon$, if for every state of the world $\omega$, each player $(1-\epsilon)$-believes in $\omega$ that some state of nature is common-$(1-\epsilon)$-belief in $\omega$.
\end{definition}

We show that when the discount factors are almost complete information with respect to $\epsilon$, according to Definition \ref{defin almost comp strong}, there is a simple conditional-grim-trigger profile, which is an $\epsilon'$-equilibrium for $\epsilon'\ge M\epsilon$ ($M>0$ is a constant which depends only on the game and not on the information structure). While this assumption can hold without the information structure derived from a common prior on the state of the world, we show that if it does, this concept of almost complete information is stronger than the concept in Definition \ref{defin almost comp MS}. Also, under this assumption (with the common prior), in the simple conditional-grim-trigger $\epsilon'$-equilibrium mentioned above the players cooperate in all the states of the world in which there is a cooperation under the conditional-grim-trigger equilibrium of maximum cooperation in the complete information case, $\eta^*(\Lambda,\Lambda)$, but for a set with small probability. In other words, under a stronger concept of almost complete information, we have a result which is quite similar to the one in Theorem B of Mondrer and Samet (1989), but with a conditional-grim-trigger profile which is defined explicitly for every state of the world.

\begin{proposition}
\label{prop almost complete}
Let $\epsilon>0$. Assume that the discount factors are almost complete information with respect to $\epsilon$, according to Definition \ref{defin almost comp strong}. Then, the strategy profile $\eta^*(B_1^{1-\epsilon}(D^{1-\epsilon}(\Lambda)),B_2^{1-\epsilon}(D^{1-\epsilon}(\Lambda)))$ is an $\epsilon'$-equilibrium, for every $\epsilon'>M\epsilon$, where $M>0$ is a constant, independent of the information structure and of $\epsilon$.
\end{proposition}

\noindent\textbf{Proof: }
Denote $K_i:=B_i^{1-\epsilon}(D^{1-\epsilon}(\Lambda))$. Note that $K_i\subseteq\Lambda_i$ for $i=1,2$. From Lemma \ref{epsilonprop}, there is an $M>0$ such that for every $\epsilon'\ge M\epsilon$ and $i=1,2$, we have $f_i^{\epsilon'}(\omega)<1-\epsilon$ for every $\omega\in K_i$ and $g_i^{\epsilon'}(\omega)>\epsilon$ for every $\omega\notin K_i$.

For every $\omega\in K_i$, we have $P_i(K_j\mid\omega)\ge P_i(D^{1-\epsilon}(\Lambda)\mid\omega)\ge1-\epsilon>f_i^{\epsilon'}(\omega)$.
For every $\omega\notin K_i$, we have from our assumption that player $i$ $(1-\epsilon)$-believes in $\omega$ that some state of nature is a common-$(1-\epsilon)$-belief in $\omega$, but this state of nature is not in $\Lambda$. Therefore, $\omega\in B_i^{1-\epsilon}(D^{1-\epsilon}(\Lambda^c))$. Therefore we have
$$
\begin{array}{l}
P_i(K_j\mid\omega)=1-P_i(K_j^c\mid\omega)=1-P_i(B_j^{1-\epsilon}(D^{1-\epsilon}(\Lambda^c))\mid\omega)\le
\\
\le1-P_i(D^{1-\epsilon}(\Lambda^c)\mid\omega)<\epsilon<g_i^{\epsilon'}(\omega).
\end{array}
$$
From Theorem \ref{epsilonconditionprop} we have that $\eta^*(K_1,K_2)$ is an $\epsilon'$-equilibrium. $\Box$

\begin{remark}
\label{remark almost comp}
\begin{enumerate}
  \item Under this strategy profile, both players will cooperate if and only if $\omega\in D^{1-\epsilon}(\Lambda)$, since $B_i^{1-\epsilon}(D^{1-\epsilon}(\Lambda))\subseteq\Lambda_i$.
  \item $D^{1-\epsilon}(\Lambda)$ may be empty, and then $\eta^*(B_1^{1-\epsilon}(D^{1-\epsilon}(\Lambda)),B_2^{1-\epsilon}(D^{1-\epsilon}(\Lambda))$ a trivial profile with no cooperation.
\end{enumerate}
\end{remark}

\begin{proposition}
\label{prop stronger than MS}
Let $\epsilon>0$. Assume that (a) the information structure is derived from a common prior $P$; (b) the discount factors are almost complete information with respect to $\epsilon$, according to Definition \ref{defin almost comp strong}; and (c) the number of state of nature is finite or countable. Then the discount factors are almost complete information with respect to $\epsilon$ and $3\epsilon$ according to Definition \ref{defin almost comp MS}.
\end{proposition}

\noindent\textbf{Proof: }
Denote by $K$ the set of states of the world $\omega$ such that there is a state of nature which is a common-$(1-\epsilon)$-belief in $\omega$. From assumption (b) we have that for every state of the world $\omega$, $P_i(K\mid\omega)\ge1-\epsilon$. Therefore $P(K)\ge1-\epsilon$. For every state of nature $\lambda$, denote $G^\lambda:=\{\omega\in\Omega\mid\lambda(\omega)=\lambda\}$, and $\Omega'=\bigcup_\lambda(G^\lambda\cap D^{1-\epsilon}(G^\lambda))$. Our goal is to show that $P(\Omega')>1-3\epsilon$. This is shown in the proof of Theorem B of Mondrer and Samet (1989). $\Box$

From Propositions \ref{prop almost complete} and \ref{lemma almost MS}, Remark \ref{remark almost comp} and Lemma \ref{lemma almost MS} we conclude the following, which is the generalization of Corollary \ref{cor 2 action almost} for games with more than two actions for each player:
\begin{corollary}
\label{cor almost comp}
Let $\epsilon>0$. Assume that (a) the information structure is derived from a common prior $P$; (b) the discount factors are almost complete information with respect to $\epsilon$, according to Definition \ref{defin almost comp strong}; and (c) the number of state of nature is finite or countable. Then (A) the strategy profile $\eta^*(B_1^{1-\epsilon}(D^{1-\epsilon}(\Lambda)),B_2^{1-\epsilon}(D^{1-\epsilon}(\Lambda))$ is an $\epsilon'$-equilibrium, for every $\epsilon'>M\epsilon$, where $M>0$ is a constant, independent of the information structure and of $\epsilon$; (B) under this strategy profile, both players will cooperate if and only if $\omega\in D^{1-\epsilon}(\Lambda)$; and (C) $P(\Lambda\setminus D^{1-\epsilon}(\Lambda))<3\epsilon$.
\end{corollary}

\begin{example}
Let $G$, $S$ and $\Omega$ be as in Example \ref{ex no complete 3}. The common prior is as follows: the state of nature is chosen uniformly. In every state of nature, there is a probability $1-\epsilon$ that both the signals are correct, and the other three states of the world that correspond to this state of nature have equal probability $\epsilon/3$. For example, if $(H_1,H_2)$ was chosen, the distribution over the states of the world is
$$
\left[(1-\epsilon)(H_1,H_2,h_1,h_2),\frac{\epsilon}{3}(H_1,H_2,l_1,h_2),
\frac{\epsilon}{3}(H_1,H_2,h_1,l_2),\frac{\epsilon}{3}(H_1,H_2,l_1,l_2)\right].
$$
We now calculate the beliefs of player 1. From symmetry, the beliefs of player 2 are similar.
$$
\begin{array}{l}
P_1((H_1,\cdot,h_1,\cdot))=
\\
\left[(1-\epsilon)(H_1,H_2,h_1,h_2),\frac{\epsilon}{3}(H_1,H_2,h_1,l_2)\right.,
\left.\frac{\epsilon}{3}(H_1,L_2,h_1,h_2),\frac{\epsilon}{3}(H_1,L_2,h_1,l_2)\right],
\end{array}
$$
$$
\begin{array}{l}
P_1((L_1,\cdot,l_1,\cdot))=
\\
\;\left[(1-\epsilon)(L_1,L_2,l_1,l_2),\frac{\epsilon}{3}(L_1,L_2,l_1,h_2)\right.,
\left.\frac{\epsilon}{3}(L_1,H_2,l_1,l_2),\frac{\epsilon}{3}(L_1,H_2,l_1,h_2)\right],
\end{array}
$$
$$
\begin{array}{l}
P_1((H_1,\cdot,l_1,\cdot))=
\\
\;\left[(1-\epsilon)(H_1,L_2,l_1,h_2),\frac{\epsilon}{3}(H_1,H_2,l_1,l_2),\right.
\left.\frac{\epsilon}{3}(H_1,L_2,l_1,l_2),\frac{\epsilon}{3}(H_1,H_2,l_1,h_2)\right],
\end{array}
$$
and
$$
\begin{array}{l}
P_1((L_1,\cdot,h_1,\cdot))=
\\
\;\left[(1-\epsilon)(L_1,H_2,h_1,l_2),\frac{\epsilon}{3}(L_1,L_2,h_1,h_2),\right.
\left.\frac{\epsilon}{3}(L_1,H_2,h_1,h_2),\frac{\epsilon}{3}(L_1,L_2,h_1,l_2)\right].
\end{array}
$$
Note that indeed, for every state of the world $\omega$, each player $(1-\epsilon)$-believes in $\omega$ that some state of nature is a common-$(1-\epsilon)$-belief in $\omega$, so the discount factors are almost complete information with respect to $\epsilon$, according to definition \ref{defin almost comp strong}.

As in Example \ref{ex no complete 3}, $\Lambda_1=\{(H_1,\cdot,\cdot,\cdot)\}$ and $\Lambda_2=\{(\cdot,H_2,\cdot,\cdot)\}$, and so $\Lambda=\{(H_1,H_2,\cdot,\cdot)\}$. Therefore $D^{1-\epsilon}=\{(H_1,H_2,h_1,h_2)\}$, and so $B_1^{1-\epsilon}(D^{1-\epsilon}(\Lambda))=\{(H_1,\cdot,h_1,\cdot)\}$ and $B_2^{1-\epsilon}(D^{1-\epsilon}(\Lambda))=\{(\cdot,H_2,\cdot,h_2)\}$. Therefore, from Corollary \ref{cor almost comp} we conclude that $\eta^*(\{(H_1,\cdot,h_1,\cdot)\},\{(\cdot,H_2,\cdot,h_2)\})$ is an $\epsilon'$-equilibrium for every $\epsilon'\ge M\epsilon$. Under this strategy profile each player cooperates if both his discount factor and his signal are high, and both players cooperate if the state of the world is $(H_1,H_2,h_1,h_2)$. Indeed, as in Corollary \ref{cor almost comp}, we have that $P(\Lambda\setminus\{(H_1,H_2,h_1,h_2)\})=\epsilon<3\epsilon$.

Note that if $H_1$ and $H_2$ are high enough, we have $P_i(\Lambda\mid\omega)\ge f_i(\omega)>f_i^{\epsilon'}(\omega)$ for every $\omega\in\Lambda_i$, and therefore $B_i^f(D^f(\Lambda))=B_i^{f^{\epsilon'}}(D^{f^{\epsilon'}}(\Lambda))=\Lambda_i$ for $i=1,2$, and $D^f(\Lambda)=D^{f^{\epsilon'}}(\Lambda)=\Lambda$. This implies that $\eta^*(\Lambda_1,\Lambda_2)$ may be an equilibrium or an $\epsilon'$-equilibrium with larger cooperation events. But, if $\epsilon$ is small enough, we have $P_1(\Lambda_2\mid(L_1,H_2,h_2,l_1))>1-\epsilon>g_1^{1,\epsilon'}\ge g_1^{2,\epsilon'}((L_1,H_2,h_2,l_1))$, so the second condition of Theorem \ref{epsilonconditionprop} does not hold and $\eta^*(\Lambda_1,\Lambda_2)$ in not an $\epsilon'$-equilibrium.
\end{example}

\section{Generalizations}

In this section we generalize the results of the previous sections in several ways. First, we see that the main result (Theorem \ref{conditionprop}) holds, with some adjustments, even when each player does not know his own discount factor. Second, we generalize Theorem \ref{conditionprop} for general games with incomplete information, with respect to a course of action that is an equilibrium in every state of the world (equivalent to $\sigma^*$), and a course of action that is an equilibrium only in some states of the world (equivalent to $\tau^*$). Last, we analyze the case of repeated games with incomplete information on the discount factor with more than two players.

\subsection{The Results when One's Own Discount Factor is Unknown}
\label{secnoselfinf}

In this section we assume that each player does not know his own discount factor in every state of the world. In this case, most of the arguments in the proof of Theorem \ref{conditionprop} still hold, with the following adjustments. In order to avoid measurability problems, we assume in this section that $\Omega$ is finite or countable, $\Sigma$ is the power set of $\Omega$, and $\Sigma_i$ is generated by the ``types'' of player $i$.

When calculating the expected payoffs, we cannot take $\lambda_i$ out of the expectation, since $\lambda_i(\omega)$ is no longer known to player $i$ given $\omega$. For example, instead of:
$$
\begin{array}{ll}
\gamma_i(\eta^*(K_1,K_2)\mid\omega)=
\\
\;\;\;\;=P_i(K_j\mid\omega)\left(\frac{u_i(\tau)}{1-\lambda_i(\omega)}\right)
+(1-P_i(K_j\mid\omega))\left(u_i(\tau_i,\sigma_j)+u_i(\sigma)\frac{\lambda_i(\omega)}{1-\lambda_i(\omega)}\right)
\end{array}
$$
when $\omega\in K_i$, the expected payoff is:
$$
\begin{array}{ll}
\gamma_i(\eta^*(K_1,K_2)\mid\omega)=
\\
\;\;\;\;=P_i(K_j\mid\omega)E_i\left(\frac{u_i(\tau)}{1-\lambda_i}\mid\omega\right)+
\\
\;\;\;\;\;\;\;\;+(1-P_i(K_j\mid\omega))\left(u_i(\tau_i,\sigma_j)+u_i(\sigma)E_i\left(\frac{\lambda_i}{1-\lambda_i}\mid\omega\right)\right).
\end{array}
$$
Similarly, $\Lambda_i$ is no longer the set $\{\omega\in\Omega\mid\lambda_i(\omega)\ge\lambda_i^0\}$, which is no longer necessarily an $i$-measurable set.
\newline
Rather $\Lambda_i:=\left\{\omega\in\Omega\mid\forall\sigma_i'\:
E_i(\frac{u_i(\tau)}{1-\lambda_i}-(u_i(\sigma_i',\tau_j)+u_i(\sigma)\frac{\lambda_i}{1-\lambda_i})\mid\omega)\ge0\right\}$.
That is the set of states of the world in which player $i$ believes he cannot profit by deviating from the profile $\tau^*$.

The main adjustment in Theorem \ref{conditionprop} itself is that it is not necessary that $K_i\subseteq\Lambda_i$. Indeed, this requirement was derived from the inequality:
        $$
        \forall \sigma_i'\ne\tau_i \;\; P_i(K_j\mid\omega)\lambda_i(\omega)\left(
        \frac{u_i(\tau)}{1-\lambda_i(\omega)}-(u_i(\sigma_i',\tau_j)+u_i(\sigma)\frac{\lambda_i(\omega)}{1-\lambda_i(\omega)})
        \right)\ge 0,
        $$
        and because $\lambda_i>0$, it follows that either $P_i(K_j\mid\omega)=0$ or $\omega\in\Lambda_i$.

When one's own discount factor is unknown, this inequality becomes
        $$
        P_i(K_j\mid\omega)E_i\left(\lambda_i\left(
        \frac{u_i(\tau)}{1-\lambda_i}-(u_i(\sigma_i',\tau_j)+u_i(\sigma)\frac{\lambda_i}{1-\lambda_i})
        \right)\mid\omega\right)\ge 0,\;\; \forall \sigma_i'\ne\tau_i,
        $$
but here $E_i\left(\lambda_i\left(
        \frac{u_i(\tau)}{1-\lambda_i}-(u_i(\sigma_i',\tau_j)+u_i(\sigma)\frac{\lambda_i}{1-\lambda_i})
        \right)\mid\omega\right)\ge 0$ is weaker than $\omega\in\Lambda_i$. Therefore, there may be games such that there are events $K_i\not\subseteq\Lambda_i$ such that $\eta^*(K_1,K_2)$ is a Bayesian equilibrium, but then the conditions are different from these of Theorem \ref{conditionprop} --- if $\omega\in K_i\setminus\Lambda_i$, player $i$ may have to $h$-believe in $K_j^c$ in the state of the world $\omega$, for a certain function $h$, since he has a profitable deviation from $\tau^*$ in $\omega$. This may be in in addition to $f$-believing in $K_j$ in the state of the world $\omega$. See Example \ref{example incomplete self info}.

Under the assumption that $K_1\subseteq\Lambda_1$ for $i=1,2$, the rest of the proof does not change, and so we get the following result.

\begin{theorem}
\label{nonselfinfoconditionprop}
In the game $G$, the strategy profile $\eta^*(K_1,K_2)$, such that $K_1\subseteq\Lambda_1$ and $K_2\subseteq\Lambda_2$, is a Bayesian equilibrium, if and only if, for $i=1,2$,
\begin{enumerate}
  \item $P_i(K_j\mid\omega)\ge f_i(\omega)$ for every $\omega\in K_i$,
  \item $P_i(K_j\mid\omega)\le g_i(\omega)$ for every $\omega\notin K_i$.
\end{enumerate}
\end{theorem}

The functions $f_i$ and $g_i$ are the same as before, with the natural adjustments; for example, instead of:
     $$
\begin{array}{l}
      f_i(\omega):=\max_{\sigma_i'\in F_i}
      \frac{u_i(\sigma_i',\sigma_j)-u_i(\tau_i,\sigma_j)}{\left(\frac{u_i(\tau)}{1-\lambda_i(\omega)}-(u_i(\sigma_i',\tau_j)
      +u_i(\sigma)\frac{\lambda_i(\omega)}{1-\lambda_i(\omega)})\right)+(u_i(\sigma_i',\sigma_j)-u_i(\tau_i,\sigma_j))},
\end{array}
      $$
now we have:
$$
\begin{array}{ll}
      f_i(\omega):=\max_{\sigma_i'\in F_i}
      \frac{u_i(\sigma_i',\sigma_j)-u_i(\tau_i,\sigma_j)}{E_i\left(\frac{u_i(\tau)}{1-\lambda_i}-(u_i(\sigma_i',\tau_j)
      +u_i(\sigma)\frac{\lambda_i}{1-\lambda_i})\mid\omega\right)+(u_i(\sigma_i',\sigma_j)-u_i(\tau_i,\sigma_j))}.
\end{array}
      $$
Note also that the sets over which the maxima and minima are taken when calculating $f$ and $g$ may change according to the same adjustments.

\begin{remark}
\begin{itemize}
  \item Note that because now we assume that $K_i\subseteq\Lambda_i$ for $i=1,2$, we can omit the assumption that $\tau_i$ is not a best response to $\sigma_j$, retaining only the weaker assumption that $\tau_i$ is not in the support of $\sigma_i$.
  \item Because of that, as mentioned in Remark \ref{supportremark}(2), we may get that $f_i=0$.
        This may also be the case even if $\tau_i$ is not a best response to $\sigma_j$ (which was not the case in the original theorem, see Remark \ref{trivialityremark}(2)).
That happens if and only if the distribution $\lambda_i(\omega)$ such that
$E_i\left(\frac{1}{1-\lambda_i}\mid\omega\right)=\infty$.
  \item In this case, the functions $f_i$ and $g_i$ have a slightly different form if we take the payoffs as $\gamma_i=E_i((1-\lambda_i)\sum_{t=1}^\infty\lambda_i^{t-1}u_i^t\mid\omega)$ (with the $(1-\lambda_i)$ in the expectation). The only significant difference is that in this case the payoffs are bounded and therefore Remark \ref{trivialityremark}(2) still holds.
\end{itemize}
\end{remark}

The following example shows that the assumption $K_i\subseteq\Lambda_i$ for $i=1,2$ in Theorem \ref{nonselfinfoconditionprop} cannot be omitted --- $\eta^*(K_1,K_2)$ may be a Bayesian equilibrium without this assumption, but then the conditions are quite different.
\begin{example}
\label{example incomplete self info}
Consider the following game:
\newline \begin{picture}(290,90)(-80,-20)
\put( 15,8){$C$}
\put( 15,28){$D$}
\put( 80,50){$D$}
\put(160,50){$C$}
\put(15,-12){$N$}
\put(240,50){$N$}
\put(40,-20){\numbercellong{$-10,\cdot$}{}}
\put( 40, 0){\numbercellong{$0,4$}{}}
\put(40,20){\numbercellong{$1,1$}{}}
\put(120,-20){\numbercellong{$10,\cdot$}{}}
\put(120,0){\numbercellong{$3,3$}{}}
\put(120,20){\numbercellong{$4,0$}{}}
\put(200,-20){\numbercellong{$\cdot,\cdot$}{}}
\put(200,0){\numbercellong{$\cdot,10$}{}}
\put(200,20){\numbercellong{$\cdot,-10$}{}}
\end{picture}
\newline
(payoffs not indicated can be arbitrary).
Here $\sigma=(D,D)$, $\tau=(C,C)$.
Both players have the same discount factor, that can be either $4/5$ or $2/5$, so that $S=\{2/5,4/5\}$. Let the information structure be the following: $\Omega=S\times S_1\times S_2$, where $S_i=\{H_i,L_i,X_i\}$ are the possible signals of player $i$. Denote the signal that player $i$ receives by $s_i$. The first coordinate is the common discount factor in each state of the world, that is $\lambda((a,s_1,s_2))=a$. Each player knows only his own signal, and his beliefs are as follows: If player $i$ gets the signal $H_i$ he believes that the state of the world is $(4/5,H_1,H_2)$: the discount factor is $4/5$ and this fact is common belief. If player $i$ gets the signal $L_i$ he believes that the state of the world is $(2/5,L_1,L_2)$: the discount factor is $2/5$ and this fact is common belief. If player $i$ gets the signal $X_i$, he believes that the state of the world is $(2/5,X_i,L_j)$ with probability $p$, and $(4/5,X_i,H_j)$ with probability $1-p$, where $3/20<p<6/23$.

The best deviation from $\tau^*$ is to play $N$ in the first stage and from the second stage on to play $D$. This deviation is profitable for player $i$ if and only if $E_i\left(u_i(C,C)\frac{1}{1-\lambda}-\left(u_i(N_i,C_j)+u_i(D,D)\frac{\lambda}{1-\lambda}\right)\mid\omega\right)\ge0$, that is if $E_i\left(\frac{2\lambda}{1-\lambda}\mid\omega\right)\ge7$. Since $p>3/20$, it follows that $\Lambda_i=\{s_i=H_i\}$.

Denote $K_i=\{s_i=H_i \mbox{ or } s_i=X_i\}$. We prove that $\eta^*(K_1,K_2)$ is a Bayesian equilibrium, even though $K_i\not\subseteq\Lambda_i$. Because the game, the information structure and $\eta^*(K_1,K_2)$ are symmetric, it is sufficient to show that there is no profitable deviation for player 1.
If player 1's signal is $s_1=H_1$, he believes that it is a common belief that $\lambda=4/5$, and that player 2 plays $\tau_2^*$. Since in this case $\omega\in\Lambda_i$, there is no profitable deviation from $\tau^*$ for player 1. If player 1's signal is $s_1=L_1$, he believes that player 2 always plays $D$, and therefore player 1 cannot profit from deviating from $\eta_1^*(K_1)$ (which in this case is ``always play $D$''). If player 1's signal is $s_1=X_1$, he plays $\tau_1^*$. Similar to the proof of Theorem \ref{conditionprop}, to prove that there is no profitable deviation, two inequalities have to be satisfied:
        $$
        P_1(K_2\mid\omega)E_1\left(\lambda\left(
        \frac{3}{1-\lambda}-(u_1(\sigma_1',C)+\frac{\lambda}{1-\lambda})
        \right)\mid\omega\right)\ge 0,
        $$
and
$$
\begin{array}{ll}
P_1(K_2\mid\omega)E_1\left(\frac{3}{1-\lambda}-\left(u_1(\sigma_1',C)
      +\frac{\lambda}{1-\lambda}\right)\mid\omega\right)+\\
      \;\;\;\;+(1-P_1(K_2\mid\omega))(0-u_1(\sigma_1',D))\ge0,
\end{array}
$$
for $\sigma_1'=D$ or $\sigma_1'=N$. For $\sigma_1'=D$, the first inequality holds since $\lambda>1/3$ with probability 1 (it is the same inequality as in the prisoner's dilemma, where $\lambda_1^0=1/3$). The second inequality is equivalent to $P_1(K_2\mid\omega)(8-\frac{20}{3}p)-1\ge0$. For $\sigma_1'=N$, the first inequality holds since $p<6/23$. The second inequality is equivalent to $-P_1(K_2\mid\omega)(9+\frac{20}{3}p)+10\ge0$. Therefore, we have that $\eta^*(K_1,K_2)$ is a Bayesian equilibrium if and only if $P_1(K_2\mid\omega)(8-\frac{20}{3}p)-1\ge0$ and $10-P_1(K_2\mid\omega)(9+\frac{20}{3}p)\ge0$. The first inequality holds for high enough $P_1(K_2\mid\omega)$, which is similar to the first condition in Theorem \ref{nonselfinfoconditionprop}, but the second inequality only holds for \emph{low} enough $P_1(K_2\mid\omega)$, which is a significantly different condition (it is $h$-believing in $K_2^c$ in addition to $f$-believing in $K_2$).

To complete the proof that $\eta^*(K_1,K_2)$, observe that when $s_1=X_1$, we have $P_1(K_2\mid\omega)=1-p$. Therefore $10-P_1(K_2\mid\omega)(9+\frac{20}{3}p)\ge0$ is equivalent to $(1-p)(1-\frac{20p}{3})+10p\ge0$. Since $p<6/23$, we have $1-\frac{20p}{3}>-\frac{17}{23}$, so $(1-p)(1-\frac{20p}{3})+10p>\frac{247}{17}p-\frac{17}{23}$, and $\frac{247}{17}p-\frac{17}{23}>0$ for every $p>3/20$. Similarly, $P_1(K_2\mid\omega)(8-\frac{20}{3}p)-1\ge0$ is equivalent to $(1-p)(8-\frac{20}{3}p)-1\ge0$, which holds for every $p\le7/10$.
\end{example}

Our results regarding games with only two actions for each player does not hold in this more general case.
Lemma \ref{2actionlemma} does not hold in this case: $g_i^1=1$, $g_i^2(\omega)=g_i^3(\omega)=1$ for every $\omega\notin \Lambda_i$, and $g_i^2(\omega)=f_i(\omega)\le g_i^3(\omega)$ for every $\omega\in \Lambda_i$ still hold, but $g_i^3(\omega)$ needs not be 1 for every $\omega\notin \Lambda_i$.
Therefore, Corollary \ref{2actioncor} does not hold in this case.

\subsection{General games}

In this subsection we expand Theorem \ref{conditionprop} to a broader class of two-players Bayesian games (not only repeated games with incomplete information regarding the discount factors). While up till now we assumed that the sets of actions $(A_i)_{i=1}^2$ and the payoff functions $(u_i)_{i=1}^2$ are independent in the state of the world, now it is not the case: for every $\omega\in\Omega$, $A_i(s(\omega))$ is the set of the actions of player $i$ in the state of the world $\omega$, and $u_i(s(\omega))$ is his payoff.

Let $G=\left(N,(S,\mathcal{S}),\Pi,(A_i)_{i\in N},(u_i)_{i\in N}\right)$ be a general two-player Bayesian game. To avoid measurability problems, assume that the states of the world are finite or countable. Assume that there is a course of action profile $\sigma^*=(\sigma_1^*,\sigma_2^*)$ such that, when the information is complete, is an equilibrium for all states of nature, and that there is another course of action profile, $\tau^*=(\tau_1^*,\tau_2^*)$ which is an equilibrium in only some states of nature. Suppose that the supports of $\tau_i^*$ and $\sigma_i^*$ are disjoint in all states of the world, that is, it is discernable whether player $i$ plays $\sigma_i^*$ or $\tau_i^*$. A strategy of player $i$ is an $i$-measurable function that assigns each state of the world a course of action of player $i$.

Let $\Lambda_i\subseteq\Omega$ be the event ``player $i$ believes that he cannot benefit by deviating from the profile $\tau^*$''. That is, for every $\omega\in\Lambda_i$ and every course of action $\sigma_i'$ of player $i$, $E_i(u_i(\tau^*)\mid\omega)\ge E_i(u_i(\sigma_i',\tau_j^*)\mid\omega)$.

\begin{theorem}
\label{generalconditionprop}
In the game $G$, there exist $i$-measurable functions $0\le f_i,g_i\le1$, $i=1,2$, such that if $K_1\subseteq\Lambda_1$ and $K_2\subseteq\Lambda_2$, the strategy profile $\eta^*(K_1,K_2)=(\eta_1^*(K_1),\eta_2^*(K_2))$,
is a Bayesian equilibrium, if and only if, for $i=1,2$,
\begin{enumerate}
  \item $P_i(K_j\mid\omega)\ge f_i(\omega)$ for every $\omega\in K_i$,
  \item $P_i(K_j\mid\omega)\le g_i(\omega)$ for every $\omega\notin K_i$.
\end{enumerate}
\end{theorem}

\begin{remark}
There are several differences between the results of this theorem and Theorem \ref{conditionprop}:
\begin{itemize}
  \item Here we assume that $K_1\subseteq\Lambda_1$ and $K_2\subseteq\Lambda_2$ whereas in Theorem \ref{conditionprop} we showed it was necessary for $\eta^*(K_1,K_2)$ to be a Bayesian equilibrium. This weakened result allows us to drop the assumption that $\tau_i^*$ is not a best response to $\sigma_j^*$, and only requires that it is discernable whether player $i$ plays $\sigma_i^*$ or $\tau_i^*$ (disjoint supports).\footnote{Still, as before, if we do assume that $\tau_i^*$ is not a best response to $\sigma_j^*$, we get $f_i>0$ , assuming that the payoffs in $G$ are bounded.}
  \item We do not assume that the realized payoffs are observed, so that all a player knows is his an expected payoff based on his information on the states of nature.
  \item We do not assume that the payoffs when $\tau^*$ is played are higher than when $\sigma^*$ is played.
\end{itemize}
\end{remark}

\noindent\textbf{Proof of Theorem \ref{generalconditionprop}: }
Similar to the proof of Theorem \ref{conditionprop}, we have two cases.

\textbf{Case 1}: $\omega\in K_i$.

The payoff of player $i$ when $\eta^*(K_1,K_2)$ is played is
$$
\begin{array}{l}
\gamma_i(\eta^*(K_1,K_2)\mid\omega)=
\\
\;\;\;\;=P_i(K_j\mid\omega)E_i(u_i(\tau^*)\mid\omega)+(1-P_i(K_j\mid\omega))E_i(u_i(\tau_i^*,\sigma_j^*)\mid\omega).
\end{array}
$$
If player $i$ deviates to a course of action $\sigma_i'\ne\tau_i^*$, then his payoff is
$$
\begin{array}{l}
\gamma_i(\sigma_i',\eta_j^*(K_2)\mid\omega)=
\\
\;\;\;\;=P_i(K_j\mid\omega)E_i(u_i(\sigma_i',\tau_j^*)\mid\omega)+
(1-P_i(K_j\mid\omega))E_i(u_i(\sigma_i',\sigma_j^*)\mid\omega).
\end{array}
$$
Because $\eta^*(K_1,K_2)$ is a Bayesian equilibrium, $\gamma_i(\eta^*)\ge\gamma_i(\sigma_i',\eta_j^*)$ for every $\sigma_i'$, or
\begin{eqnarray}
\label{eq5}
P_i(K_j\mid\omega)\big(E_i(u_i(\tau^*)-u_i(\sigma_i',\tau_j^*)\mid\omega)-E_i(u_i(\tau_i^*,\sigma_j^*)-u_i(\sigma_i',\sigma_j^*)\mid\omega)\big)+
\nonumber\\
+E_i(u_i(\tau_i^*,\sigma_j^*)-u_i(\sigma_i',\sigma_j^*)\mid\omega)\ge0,
\end{eqnarray}
for every $\sigma_i'\ne\tau_i^*$.
By assumption $K_i\subseteq\Lambda_i$, so that $E_i(u_i(\tau^*)-u_i(\sigma_i',\tau_j^*)\mid\omega)\ge0$. If $E_i(u_i(\tau_i^*,\sigma_j^*)-u_i(\sigma_i',\sigma_j^*)\mid\omega)\ge0$, then inequality (\ref{eq5}) trivially holds. Otherwise, inequality (\ref{eq5}) is equivalent to $P_i(K_j\mid\omega)\ge f_i(\omega)$, where
$$
f_i(\omega):=\sup_{\sigma_i'\in F_i}
\frac{E_i(u_i(\sigma_i',\sigma_j^*)-u_i(\tau_i^*,\sigma_j^*)\mid\omega)}
{E_i(u_i(\sigma_i',\sigma_j^*)-u_i(\tau_i^*,\sigma_j^*)\mid\omega)+E_i(u_i(\tau^*)-u_i(\sigma_i',\tau_j^*)\mid\omega)},
$$
and $F_i=\{\sigma_i'\mid E_i(u_i(\tau_i^*,\sigma_j^*)\mid\omega)<E_i(u_i(\sigma_i',\sigma_j^*)\mid\omega)\}$.

\textbf{Case 2}: $\omega\notin K_i$.

The payoff of player $i$ under $\eta^*(K_1,K_2)$ is
$$
\begin{array}{l}
\gamma_i(\eta^*(K_1,K_2)\mid\omega)=
\\
\;\;\;\;=P_i(K_j\mid\omega)E_i(u_i(\sigma_i,\tau_j)\mid\omega)+(1-P_i(K_j\mid\omega))E_i(u_i(\sigma)\mid\omega).
\end{array}
$$
If player $i$ deviates to a course of action $\sigma_i'\ne\sigma_i^*$, then his payoff is
$$
\begin{array}{ll}
\gamma_i(\sigma_i',\eta_j^*(K_j)\mid\omega)=
\\
\;\;\;\;=P_i(K_j\mid\omega)E_i(u_i(\sigma_i',\tau_j^*)\mid\omega)+
(1-P_i(K_j\mid\omega))E_i(u_i(\sigma_i',\sigma_j^*)\mid\omega).
\end{array}
$$
Since $\eta^*(K_1,K_2)$ is a Bayesian equilibrium,

\noindent $\gamma_i(\eta^*(K_1,K_2)\mid\omega)\ge\gamma_i(\sigma_i',\eta_j^*(K_j)\mid\omega)$ for every $\sigma_i'\ne\sigma_i^*$, or equivalently
\begin{eqnarray}
\label{eq6}
P_i(K_j\mid\omega)
\big(E_i(u_i(\sigma_i^*,\tau_j^*)-u_i(\sigma_i',\tau_j^*)\mid\omega)-E_i(u_i(\sigma^*)-u_i(\sigma_i',\sigma_j^*)\mid\omega)\big)+
\nonumber\\
+E_i(u_i(\sigma^*)-u_i(\sigma_i',\sigma_j^*)\mid\omega)\ge0,
\end{eqnarray}
for every $\sigma_i'\ne\sigma_i^*$. Because $\sigma^*$ is an equilibrium in every state of nature, $E_i(u_i(\sigma)-u_i(\sigma_i',\sigma_j)\ge0$. Therefore, if
$E_i(u_i(\sigma_i,\tau_j)-u_i(\sigma_i',\tau_j)\ge0$, inequality (\ref{eq6}) trivially holds. Otherwise, inequality (\ref{eq6}) is equivalent to $P_i(K_j\mid\omega)\le g_i(\omega)$ where
$$
g_i(\omega):=\inf_{\sigma_i'\in G_i}
\frac{E_i(u_i(\sigma^*)-u_i(\sigma_i',\sigma_j^*)\mid\omega)}
{E_i(u_i(\sigma^*)-u_i(\sigma_i',\sigma_j^*)\mid\omega)+E_i(u_i(\sigma_i',\tau_j^*)-u_i(\sigma_i^*,\tau_j^*)\mid\omega)},
$$
and $G_i=\{\sigma_i'\mid E_i(u_i(\sigma_i^*,\tau_j^*)\mid\omega)<E_i(u_i(\sigma_i',\tau_j^*))\mid\omega)\}$.
$\Box$

\subsection{More than Two Players}

In this section we show that while the same analysis can be used to derive necessary and sufficient conditions for conditional-grim-trigger equilibria in repeated games with incomplete information on the discount factors with more than two players, these conditions are much more complex than in two player games. We provide simple conditions, similar to those in previous sections (i.e., $f$-believing in an event at a state of the world), which are sufficient for a conditional-grim-trigger strategy profile to be a Bayesian equilibrium, but are \emph{not} necessary (Theorem \ref{theorem more than 2 players}).

Let $\Gamma=\left(N,(A_i)_{i\in N},(u_i)_{i\in N}\right)$ be a one-shot game, with $N=\{1,2,...,N\}$ the set of players.
Let $\sigma=(\sigma_1,\sigma_2,...,\sigma_N)$ be a mixed-strategies Nash equilibrium in $\Gamma$.
Assume that the payments in another non-equilibrium pure action profile $\tau=(\tau_1,\tau_2,...,\tau_N)$, are higher than the equilibrium payments, that is $u_i(\tau)>u_i(\sigma)$ for every $i\in N$.
Also, assume that for every $i\in N$, $\tau_i$ is not in the support of $\sigma_i$.

Let $G=\left(N,(S,\mathcal{S}),\Pi,(A_i)_{i\in N},(u_i)_{i\in N}\right)$ be the repeated game based on $\Gamma$, with incomplete information regarding the discount factors, similar to the one described in Section \ref{section model}, where each player knows his own discount factor.

Let $\tau_i^*$ and $\sigma_i^*$ be defined as in the two-player case.\footnote{When there are more than two players, $\tau_i^*$ is triggered whenever there is \emph{at least one} player that deviate from $\tau^*$.} Again, $\sigma^*$ is an equilibrium course of action regardless of the discount factors. Player $i$ does not have a profitable deviation from $\tau^*$ if and only if $\lambda_i>\lambda_i^0$, where
$$\lambda_i^0:=\min\left\{\lambda_i\mid
\frac{u_i(\tau)}{1-\lambda_i}-(u_i(\sigma_i',\tau_{-i})+u_i(\sigma)\frac{\lambda_i}{1-\lambda_i})\ge0 \;\;\forall \sigma_i'\ne\tau_i\right\}.$$
As before, denote $\Lambda_i:=\{\omega\in\Omega\mid\lambda_i(\omega)\ge\lambda_i^0\}$.

The conditional-grim-trigger strategy is defined in the same way as in the two-player case, only now there is a cooperation event for eack player in $N$, so the strategy profile is $\eta^*(K_1,K_2,...,K_N)=(\eta_1^*(K_1),\eta_2^*(K_2),...,\eta_N^*(K_N))$. The following theorem gives sufficient conditions that guarantee that the profile $\eta^*(K_1,K_2,...,K_N)$ is a Bayesian equilibrium.

\begin{theorem}
\label{theorem more than 2 players}
Suppose that $K_i\subseteq\Lambda_i$, for every $i\in N$. The strategy profile $\eta^*(K_1,K_2,...,K_N)$ is a Bayesian equilibrium in the game $G$, if, for every $i\in N$,
\begin{enumerate}
  \item $P_i(\bigcap_{j\ne i}K_j\mid\omega)\ge f_i(\omega)$ for every $\omega\in K_i$,
  \item $P_i(\bigcup_{j\ne i}K_j\mid\omega)\le g_i(\omega)$ for every $\omega\notin K_i$.
\end{enumerate}
Where the functions $f_i$ and $g_i$ are defined by:
$$
f_i(\omega):=
      \max_{\sigma_i'\ne\tau_i}
      \frac{2M_i}{\left(\frac{u_i(\tau)}{1-\lambda_i(\omega)}-(u_i(\sigma_i',\tau_{-i})
      +u_i(\sigma)\frac{\lambda_i(\omega)}{1-\lambda_i(\omega)})\right)+2M_i},
$$
and $g_i(\omega):=\min\{g_i^1,g_i^2(\omega)\}$, where
$$
g_i^1:=
      \max_{\sigma_i'\ne\tau_i,\sigma_i}
      \frac{u_i(\sigma)-u_i(\sigma_i',\sigma_{-i})}{u_i(\sigma)-u_i(\sigma_i',\sigma_{-i})+2M_i},
$$
$$
g_i^2(\omega):=
      \frac{(1-\lambda_i(\omega))(u_i(\sigma)-u_i(\tau_i,\sigma_{-i}))}
      {(1-\lambda_i(\omega))(u_i(\sigma)-u_i(\tau_i,\sigma_{-i}))+2M_i},
$$
and $M_i:=\max\{|u_i(a_1,...,a_N)|\mid a_j\in A_j \mbox{ for } j\in N\}$.
\end{theorem}

These conditions mean that in order for $\eta^*(K_1,K_2,...,K_N)$ to be a Bayesian equilibrium, player $i$ needs to $f$-believe in a the event $\bigcap_{j\ne i}K_j$ whenever he plays $\tau_i^*$, and to $(1-g)$-believe in the event $\bigcup_{j\ne i}K_j$, whenever he plays $\sigma_i^*$. Note that when there are more then two players, these two events may not be the same.

\noindent\textbf{Proof: }The proof follows the same lines as the proof of Theorem \ref{conditionprop}. For simplicity, we assume that $N=3$. The analysis for games with more than three players is similar. Without loss of generality, it is sufficient to prove that player 1 does not have a profitable deviation.

\noindent \textbf{Case 1:} $\omega\in K_1$.

Player 1's payoff under the strategy profile $\eta^*(K_1,K_2,K_3)$ is
$$
\begin{array}{ll}
\gamma_1(\eta^*(K_1,K_2,K_3)\mid\omega)=
\\
\;\;\;\;P_1(K_2\cap K_3\mid\omega)\left(\frac{u_1(\tau)}{1-\lambda_1(\omega)}\right)+
\\
\;\;\;\;P_1(K_2\setminus K_3\mid\omega)\left(u_1(\tau_1,\tau_2,\sigma_3)+u_1(\sigma)\frac{\lambda_1(\omega)}{1-\lambda_1(\omega)}\right)+
\\
\;\;\;\;P_1(K_3\setminus K_2\mid\omega)\left(u_1(\tau_1,\sigma_2,\tau_3)+u_1(\sigma)\frac{\lambda_1(\omega)}{1-\lambda_1(\omega)}\right)+
\\
\;\;\;\;(1-P_1(K_2\cup K_3\mid\omega)) \left(u_1(\tau_1,\sigma_2,\sigma_3)+u_1(\sigma)\frac{\lambda_1(\omega)}{1-\lambda_1(\omega)}\right).
\end{array}
$$
As in the proof of Theorem \ref{conditionprop}, we need to consider the following deviations:
\begin{itemize}
\item Let $\sigma_1'\ne\tau_1$, and define the course of action $\sigma_1'^{**}$ by: Player 1 plays $\tau_1$ in the first stage, an if the action profile $\tau$ was played in the first stage, player 1 plays a pure action $\sigma_1'$ in stage $2$ and $\sigma_1$ afterwards. If the action profile $\tau$ was not played in the first stage, player 1 plays $\sigma_1$ from the second stage onwards.
If player 1 plays $\sigma_1'^{**}$, his payoff is:
$$
\begin{array}{ll}
\gamma_1(\sigma_1'^{**},\eta_2^*(K_2),\eta_3^*(K_3)\mid\omega)=
\\
\;\;\;\;P_1(K_2\cap K_3\mid\omega)\left(u_1(\tau)+u_1(\sigma_1',\tau_2,\tau_3)\lambda_1(\omega)
+u_1(\sigma)\frac{(\lambda_1(\omega))^2}{1-\lambda_1(\omega)}\right)+
\\
\;\;\;\;P_1(K_2\setminus K_3\mid\omega)\left(u_1(\tau_1,\tau_2,\sigma_3)+u_1(\sigma)\frac{\lambda_1(\omega)}{1-\lambda_1(\omega)}\right)+
\\
\;\;\;\;P_1(K_3\setminus K_2\mid\omega)\left(u_1(\tau_1,\sigma_2,\tau_3)+u_1(\sigma)\frac{\lambda_1(\omega)}{1-\lambda_1(\omega)}\right)+
\\
\;\;\;\;(1-P_1(K_2\cup K_3\mid\omega)) \left(u_1(\tau_1,\sigma_2,\sigma_3)+u_1(\sigma)\frac{\lambda_1(\omega)}{1-\lambda_1(\omega)}\right).
\end{array}
$$
Since $\eta^*(K_1,K_2,K_3)$ is a Bayesian equilibrium, $\gamma_1(\eta^*(K_1,K_2,K_3)\mid\omega)\ge\gamma_1(\sigma_1'^{**},\eta_2^*(K_2),\eta_3^*(K_3)\mid\omega)$, or equivalently,
$$
\begin{array}{l}
P_1(K_2\cap K_3\mid\omega)\lambda_1(\omega)\left(
\frac{u_1(\tau)}{1-\lambda_1(\omega)}-(u_1(\sigma_1',\tau_2,\tau_3)+u_1(\sigma)\frac{\lambda_1(\omega)}{1-\lambda_1(\omega)})
\right)\ge 0,
\end{array}
$$
for every $\sigma_1'\ne\tau_1$.
Because $\lambda_1>0$, either $P_1(K_2\cap K_3\mid\omega)=0$ or $\omega\in\Lambda_1$, which is a similar result to the two-player case.
\item Let $\sigma_1'\ne\tau_1$, and define the course of action $\sigma_1'^*$ by: Player 1 plays $\sigma_1'$ in the first stage and $\sigma_1$ afterwards.
If player 1 plays $\sigma_1'^*$, his payoff is:
$$
\begin{array}{ll}
\gamma_1(\sigma_1'^*,\eta_2^*(K_2),\eta_3^*(K_3)\mid\omega)=
\\
\;\;\;\;P_1(K_2\cap K_3\mid\omega)u_1(\sigma_1',\tau_2,\tau_3)+
\;\;\;\;P_1(K_2\setminus K_3\mid\omega)u_1(\sigma_1',\tau_2,\sigma_3)+
\\
\;\;\;\;P_1(K_3\setminus K_2\mid\omega)u_1(\sigma_1',\sigma_2,\tau_3+
(1-P_1(K_2\cup K_3\mid\omega))u_1(\sigma_1',\sigma_2,\sigma_3)+
\\
\;\;\;\;+u_1(\sigma)\frac{\lambda_1(\omega)}{1-\lambda_1(\omega)}.
\end{array}
$$
Since $\eta^*(K_1,K_2,K_3)$ is a Bayesian equilibrium, $\gamma_1(\eta^*(K_1,K_2,K_3)\mid\omega)\ge\gamma_1(\sigma_1'^*,\eta_2^*(K_2),\eta_3^*(K_3)\mid\omega)$, or equivalently,
\begin{eqnarray}
\label{eq7}
P_1(K_2\cap K_3\mid\omega)\left(\frac{u_1(\tau)}{1-\lambda_1(\omega)}-
(u_1(\sigma_1',\tau_2,\tau_3)+u_1(\sigma)\frac{\lambda_1(\omega)}{1-\lambda_1(\omega)})\right)+
\nonumber\\
P_1(K_2\setminus K_3\mid\omega)(u_1(\tau_1,\tau_2,\sigma_3)-u_1(\sigma_1',\tau_2,\sigma_3))+
\nonumber\\
P_1(K_3\setminus K_2\mid\omega)(u_1(\tau_1,\sigma_2,\tau_3)-u_1(\sigma_1',\sigma_2,\sigma_3))+
\nonumber\\
(1-P_1(K_2\cup K_3\mid\omega)) (u_1(\tau_1,\sigma_2,\sigma_3)-u_1(\sigma_1',\sigma_2,\sigma_3))\ge0.
\end{eqnarray}
This is the analogue of inequality (\ref{eq1}) in the two-players game, only here this condition is \emph{not} equivalent to a simple $f$-belief type condition.
\newline
However, the left hand side of inequality (\ref{eq7}) is no less than
$$
\begin{array}{l}
P_1(K_2\cap K_3\mid\omega)\left(\frac{u_1(\tau)}{1-\lambda_1(\omega)}-
(u_1(\sigma_1',\tau_2,\tau_3)+u_1(\sigma)\frac{\lambda_1(\omega)}{1-\lambda_1(\omega)})\right)-
\\
2M_1(1-P_1(K_2\cap K_3\mid\omega)),
\end{array}
$$
and so it is a sufficient condition that
$$
\begin{array}{l}
P_1(K_2\cap K_3\mid\omega)\left(\frac{u_1(\tau)}{1-\lambda_1(\omega)}-
(u_1(\sigma_1',\tau_2,\tau_3)+u_1(\sigma)\frac{\lambda_1(\omega)}{1-\lambda_1(\omega)})\right)-
\\
2M_1(1-P_1(K_2\cap K_3\mid\omega))\ge0,
\end{array}
$$
for every $\sigma_1'\ne\tau_1$, which is equivalent to $K_1\subset\Lambda_1$ and $P_1(K_2\cap K_3\mid\omega)\ge f_i(\omega)$.
\end{itemize}

\noindent \textbf{Case 2:} $\omega\notin K_1$.

Player 1's payoff under the strategy profile $\eta^*(K_1,K_2,K_3)$ is
$$
\begin{array}{ll}
\gamma_1(\eta^*(K_1,K_2,K_3)\mid\omega)=
\\
\;\;\;\;P_1(K_2\cap K_3\mid\omega)\left(u_1(\sigma_1,\tau_2,\tau_3)+u_1(\sigma)\frac{\lambda_1(\omega)}{1-\lambda_1(\omega)}\right)+
\\
\;\;\;\;P_1(K_2\setminus K_3\mid\omega)\left(u_1(\sigma_1,\tau_2,\sigma_3)+u_1(\sigma)\frac{\lambda_1(\omega)}{1-\lambda_1(\omega)}\right)+
\\
\;\;\;\;P_1(K_3\setminus K_2\mid\omega)\left(u_1(\sigma_1,\sigma_2,\tau_3)+u_1(\sigma)\frac{\lambda_1(\omega)}{1-\lambda_1(\omega)}\right)+
\\
\;\;\;\;(1-P_1(K_2\cup K_3\mid\omega))\frac{u_1(\sigma)}{1-\lambda_1(\omega)}.
\end{array}
$$
As in the proof of Theorem \ref{conditionprop}, we need to consider the following deviations:
\begin{itemize}
  \item Deviation to $\sigma_1'^*$, for $\sigma_1'\ne\tau_1,\sigma_1$ ($\sigma_1'$ may be in the support of $\sigma_1$, if $\sigma_1$ is not a pure action). The payoff is:
$$
\begin{array}{ll}
\gamma_1(\sigma_1'^*,\eta_2^*(K_2),\eta_3^*(K_3)\mid\omega)=
\\
\;\;\;\;P_1(K_2\cap K_3\mid\omega)u_1(\sigma_1',\tau_2,\tau_3)+
P_1(K_2\setminus K_3\mid\omega)u_1(\sigma_1',\tau_2,\sigma_3)+
\\
\;\;\;\;P_1(K_3\setminus K_2\mid\omega)u_1(\sigma_1',\sigma_2,\tau_3+
(1-P_1(K_2\cup K_3\mid\omega))u_1(\sigma_1',\sigma_2,\sigma_3)+
\\
\;\;\;\;+u_1(\sigma)\frac{\lambda_1(\omega)}{1-\lambda_1(\omega)}.
\end{array}
$$
Since $\eta^*(K_1,K_2,K_3)$ is a Bayesian equilibrium, $\gamma_1(\eta^*(K_1,K_2,K_3)\mid\omega)\ge\gamma_1(\sigma_1'^*,\eta_2^*(K_2),\eta_3^*(K_3)\mid\omega)$, or equivalently,
\begin{eqnarray}
\label{eq8}
P_1(K_2\cap K_3\mid\omega)(u_1(\sigma_1,\tau_2,\tau_3)-u_1(\sigma_1',\tau_2,\tau_3))+
\nonumber\\
P_1(K_2\setminus K_3\mid\omega)(u_1(\sigma_1,\tau_2,\sigma_3)-u_1(\sigma_1',\tau_2,\sigma_3))+
\nonumber\\
P_1(K_3\setminus K_2\mid\omega)(u_1(\sigma_1,\sigma_2,\tau_3)-u_1(\sigma_1',\sigma_2,\sigma_3))+
\nonumber\\
(1-P_1(K_2\cup K_3\mid\omega)) (u_1(\sigma_2)-u_1(\sigma_1',\sigma_2,\sigma_3))\ge0.
\end{eqnarray}
This is the analogue of inequality (\ref{eq2}) in the two-players game, only here this condition is \emph{not} equivalent to a simple $f$-belief type condition.
\newline
However, the left hand side of inequality (\ref{eq8}) is no less than
$$
\begin{array}{l}
(1-P_1(K_2\cup K_3\mid\omega))(u_1(\sigma)-u_1(\sigma_1',\sigma_2,\sigma_3))-
2M_1P_1(K_2\cup K_3\mid\omega),
\end{array}
$$
and so it is a sufficient condition that
$$
\begin{array}{l}
(1-P_1(K_2\cup K_3\mid\omega))(u_1(\sigma)-u_1(\sigma_1',\sigma_2,\sigma_3))-
\\
2M_1P_1(K_2\cup K_3\mid\omega)\ge0,
\end{array}
$$
for every $\sigma_1'\ne\tau_1,\sigma_1$, which is equivalent to $P_1(K_2\cup K_3\mid\omega)\le g_i^1$.
  \item Deviation to $\tau_1^*$. The payoff is:
$$
\begin{array}{ll}
\gamma_1(\tau_1^*,\eta_2^*(K_2),\eta_3^*(K_3)\mid\omega)=
\\
\;\;\;\;P_1(K_2\cap K_3\mid\omega)\left(\frac{u_1(\tau)}{1-\lambda_1(\omega)}\right)+
\\
\;\;\;\;P_1(K_2\setminus K_3\mid\omega)\left(u_1(\tau_1,\tau_2,\sigma_3)+u_1(\sigma)\frac{\lambda_1(\omega)}{1-\lambda_1(\omega)}\right)+
\\
\;\;\;\;P_1(K_3\setminus K_2\mid\omega)\left(u_1(\tau_1,\sigma_2,\tau_3)+u_1(\sigma)\frac{\lambda_1(\omega)}{1-\lambda_1(\omega)}\right)+
\\
\;\;\;\;(1-P_1(K_2\cup K_3\mid\omega)) \left(u_1(\tau_1,\sigma_2,\sigma_3)+u_1(\sigma)\frac{\lambda_1(\omega)}{1-\lambda_1(\omega)}\right).
\end{array}
$$
Since $\eta^*(K_1,K_2,K_3)$ is a Bayesian equilibrium, $\gamma_1(\eta^*(K_1,K_2,K_3)\mid\omega)\ge\gamma_1(\tau_1^*,\eta_2^*(K_2),\eta_3^*(K_3)\mid\omega)$, or equivalently,
\begin{eqnarray}
\label{eq9}
P_1(K_2\cap K_3\mid\omega)\left(u_1(\sigma_1,\tau_2,\tau_3)+u_1(\sigma)\frac{\lambda_1(\omega)}{1-\lambda_1(\omega)}
-\frac{u_1(\tau)}{1-\lambda_1(\omega)}\right)+
\nonumber\\
P_1(K_2\setminus K_3\mid\omega)(u_1(\sigma_1,\tau_2,\sigma_3)-u_1(\tau_1,\tau_2,\sigma_3))+
\nonumber\\
P_1(K_3\setminus K_2\mid\omega)(u_1(\sigma_1,\sigma_2,\tau_3)-u_1(\tau_1,\sigma_2,\sigma_3))+
\nonumber\\
(1-P_1(K_2\cup K_3\mid\omega)) (u_1(\sigma)-u_1(\tau_1,\sigma_2,\sigma_3))\ge0.
\end{eqnarray}
This is the analogue of inequality (\ref{eq3}) in the two-players game, only here this condition is \emph{not} equivalent to a simple $f$-belief type condition.
\newline
However, the left hand side of inequality (\ref{eq9}), multiplied by $1-\lambda_1(\omega)$, is no less than
$$
\begin{array}{l}
(1-P_1(K_2\cup K_3\mid\omega))(u_1(\sigma)-u_1(\tau_1,\sigma_2,\sigma_3))(1-\lambda_1(\omega))-
\\
2M_1P_1(K_2\cup K_3\mid\omega),
\end{array}
$$
and so it is a sufficient condition that
$$
\begin{array}{l}
(1-P_1(K_2\cup K_3\mid\omega))(u_1(\sigma)-u_1(\tau_1,\sigma_2,\sigma_3))(1-\lambda_1(\omega))-
\\
2M_1P_1(K_2\cup K_3\mid\omega)\ge0,
\end{array}
$$
which is equivalent to $P_1(K_2\cup K_3\mid\omega)\le g_i^2(\omega)$.
  \item Deviation to $\sigma_1'^1$, defined by:
        \begin{itemize}
          \item Play $\tau_1$ in the first stage.
          \item If the profile $\tau$ was played in the first stage, play a pure $\sigma_1'\ne\tau_1$, and afterwards $\sigma_1$.
          \item If the profile $\tau$ was not played in the first stage, play $\sigma_1$ from the second stage onwards.
        \end{itemize}
        The payoff is:
$$
\begin{array}{ll}
\gamma_1(\sigma_1'^1,\eta_2^*(K_2),\eta_3^*(K_3)\mid\omega)=
\\
\;\;\;\;P_1(K_2\cap K_3\mid\omega)\left(u_1(\tau)+u_1(\sigma_1',\tau_2,\tau_3)\lambda_1(\omega)+
u_i(\sigma)\frac{(\lambda_1(\omega))^2}{1-\lambda_1(\omega)}\right)+
\\
\;\;\;\;P_1(K_2\setminus K_3\mid\omega)\left(u_1(\tau_1,\tau_2,\sigma_3)+u_1(\sigma)\frac{\lambda_1(\omega)}{1-\lambda_1(\omega)}\right)+
\\
\;\;\;\;P_1(K_3\setminus K_2\mid\omega)\left(u_1(\tau_1,\sigma_2,\tau_3)+u_1(\sigma)\frac{\lambda_1(\omega)}{1-\lambda_1(\omega)}\right)+
\\
\;\;\;\;(1-P_1(K_2\cup K_3\mid\omega)) \left(u_1(\tau_1,\sigma_2,\sigma_3)+u_1(\sigma)\frac{\lambda_1(\omega)}{1-\lambda_1(\omega)}\right).
\end{array}
$$
Since $\eta^*(K_1,K_2,K_3)$ is a Bayesian equilibrium, $\gamma_1(\eta^*(K_1,K_2,K_3)\mid\omega)\ge\gamma_1(\sigma_1'^1,\eta_2^*(K_2),\eta_3^*(K_3)\mid\omega)$, or equivalently,
\begin{eqnarray}
\label{eq10}
P_1(K_2\cap K_3\mid\omega)\left(u_1(\sigma_1,\tau_{-1})-u_1(\tau)+(u_1(\sigma)-u_1(\sigma_1',\tau_{-1}))\lambda_1(\omega)\right)+
\nonumber\\
P_1(K_2\setminus K_3\mid\omega)(u_1(\sigma_1,\tau_2,\sigma_3)-u_1(\tau_1,\tau_2,\sigma_3))+
\nonumber\\
P_1(K_3\setminus K_2\mid\omega)(u_1(\sigma_1,\sigma_2,\tau_3)-u_1(\tau_1,\sigma_2,\sigma_3))+
\nonumber\\
(1-P_1(K_2\cup K_3\mid\omega)) (u_1(\sigma)-u_1(\tau_1,\sigma_2,\sigma_3))\ge0.
\end{eqnarray}
This is the analogue of inequality (\ref{eq4}) in the two-players game, only here this condition is \emph{not} equivalent to a simple $f$-belief type condition.
\newline
However, the left hand side of inequality (\ref{eq10}) is no less than
$$
\begin{array}{l}
(1-P_1(K_2\cup K_3\mid\omega))(u_1(\sigma)-u_1(\tau_1,\sigma_2,\sigma_3))-
\\
2M_1P_1(K_2\cup K_3\mid\omega)(1+\lambda_1(\omega)),
\end{array}
$$
and so it is a sufficient condition that
$$
\begin{array}{l}
(1-P_1(K_2\cup K_3\mid\omega))(u_1(\sigma)-u_1(\tau_1,\sigma_2,\sigma_3))-
\\
2M_1P_1(K_2\cup K_3\mid\omega)(1+\lambda_1(\omega))\ge0,
\end{array}
$$
which is a weaker condition than $P_1(K_2\cup K_3\mid\omega)\le g_i^2(\omega)$.
\end{itemize}
The proof that $f_i$ and $g_i$ are $i$-measurable is the same as in the proof of Theorem \ref{conditionprop}. $\Box$

%

\end{document}